%% file: paper_arxiv.tex
\documentclass[preprint,12pt]{elsarticle}

\usepackage{amsmath}
\usepackage{amssymb}
\usepackage{amsthm}
\usepackage{algorithm}
\usepackage{algpseudocode}
\usepackage{tikz}
\usepackage{pgfplots}
\pgfplotsset{compat=1.18}
\usetikzlibrary{arrows.meta,positioning,shapes.geometric,fit,backgrounds,calc,patterns}
\usepgfplotslibrary{groupplots}
\usepackage{booktabs}
\usepackage{stfloats}
\usepackage{listings}
\usepackage{makecell}
\usepackage[colorlinks=true,allcolors=blue!60!black]{hyperref}
\usepackage{microtype}

\newif\ifpaperarxiv
\paperarxivtrue
\ifpaperarxiv
  \newenvironment{widefigure}{\begin{figure}}{\end{figure}}
\else
  \newenvironment{widefigure}{\begin{figure*}}{\end{figure*}}
\fi
\newcommand{\benchwidth}{\linewidth}
\newcommand{\defaultstablewidth}{\textwidth}
\newenvironment{inputlisting}{}{}

\newcommand{\Depth}{\tau}
\newcommand{\Rank}{\rho}
\newcommand{\Pop}{\omega}

\theoremstyle{plain}

\lstdefinestyle{yaml}{
  basicstyle=\ttfamily\scriptsize,
  breaklines=true,
  columns=fullflexible,
  keepspaces=true,
  showstringspaces=false,
  commentstyle=\itshape\color{gray},
  numbers=left,
  numberstyle=\tiny\color{gray},
  numbersep=4pt,
  frame=single,
  framesep=2pt,
  xleftmargin=10pt,
  literate={-}{-}1,
}
\newcommand{\psentry}[2]{\par\medskip\noindent\textbf{#1:} #2}
\newcommand{\psref}[2]{\par\addvspace{2pt}\emergencystretch=2em\noindent\hangindent=1.8em\hangafter=1[#1]~#2}
\newenvironment{programsummary}{%
  \par\medskip\noindent\rule{\columnwidth}{0.4pt}\par\smallskip
  \noindent{\large\bfseries PROGRAM SUMMARY}\par\smallskip
  \noindent\rule{\columnwidth}{0.4pt}\par
  \small\sloppy
}{%
  \par\smallskip\noindent\rule{\columnwidth}{0.4pt}\par\medskip
  \normalsize
}

\begin{document}

\begin{frontmatter}

\title{SecRed: A program for sector-wise IBP reduction of Feynman integrals}

\author[ihep,ucas]{Ruichang Niu\corref{cor1}}
\ead{niuruichang@ihep.ac.cn}
\author[ihep,ucas]{Chao Zeng}
\author[ihep,ucas]{Bin Gong}
\author[ihep,ucas]{Jianxiong Wang}

\cortext[cor1]{Corresponding author.}

\address[ihep]{Institute of High Energy Physics, Chinese Academy of Sciences, Beijing 100049, China}
\address[ucas]{University of Chinese Academy of Sciences, Beijing 100049, China}

\begin{abstract}
SecRed is a C++ program for the reduction of multi-loop Feynman integrals
with integration-by-parts (IBP) identities. It reduces an integral family sector by
sector for memory control and parallelism, and adds algorithms designed for
that setting: an adaptive seed box with a corner cut, an optional seed
truncation, a subsector level filter, and symmetry relations expanded
only from the master integrals of a sector. A global stage assembles the
retained equations and performs the finite-field reconstruction of the
coefficients using FiniteFlow or FireFly, repeating the elimination over a
sequence of primes. On the
elimination side, a coefficient-free plan records the elimination order once
and is replayed at each prime, so that only the arithmetic is repeated, and a
head score orders the equations to reduce fill-in during the elimination.
The program is demonstrated on a two-loop five-point non-planar family and on
a four-loop cut family, where assembling the global equation system as a whole
would exceed the memory of a single node.
\end{abstract}

\begin{keyword}
Feynman integrals \sep Integration-by-parts reduction \sep Laporta algorithm
\sep finite-field methods \sep parallel computing
\end{keyword}

\end{frontmatter}

\newpage
\begin{programsummary}
\psentry{Program Title}{SecRed 1.0}
\psentry{CPC Library link to program files}{to be added by Technical Editor}
\psentry{Developer's repository link}{\href{https://github.com/chiyutuci/secred}{https://github.com/chiyutuci/secred}}
\psentry{Licensing provisions}{GPL-3.0}
\psentry{Programming language}{C++}
\psentry{External routines/libraries}{FLINT~\cite{flint}
(arbitrary-precision arithmetic), GiNaC~\cite{Bauer:2000cp} (symbolic
manipulation), yaml-cpp~\cite{yamlcpp} (input parsing),
FiniteFlow~\cite{Peraro:2019sve} (finite-field linear solving and
rational-function reconstruction), FireFly~\cite{Klappert:2020firefly}
(reconstruction of symbolic systems), and a C++ thread
pool~\cite{Shoshany:2024threadpool}}
\psentry{Nature of problem}{Integration-by-parts (IBP) reduction expresses a
large number of Feynman integrals of a given family as linear combinations of
a small set of master integrals. The Laporta algorithm generates IBP and
Lorentz-invariance (LI) identities for all seed integrals below prescribed
depth and rank bounds and solves the resulting linear system. For high-loop,
many-leg and multi-scale families the system grows as a power of the seed
bounds. Two problems follow. First, the elimination has to be repeated at
every prime of the coefficient reconstruction, and each pass becomes slow and
time-consuming as the system grows. Second, the complete set of raw equations
can exceed the memory of a single machine, so the system cannot even be
assembled as a whole.}
\psentry{Solution method}{The reduction is carried out sector by sector. Each
sector receives the targets that higher sectors and the user require, derives
an adaptive seed box from them, cuts the upper-right corner of the depth--rank
rectangle, can truncate the seeds that are expected to be intermediate
integrals, and, when enabled, widens the truncation if the settled basis shows
that it cut too deeply.
It also filters subsector columns far below the box top. The retained
system is solved over a finite field. Only the linearly independent equations
and the local masters are kept. The global stage assembles these equations
and reconstructs the coefficients with FiniteFlow for numeric systems or
FireFly for symbolic ones. Sectors run in parallel under a memory-aware
scheduler that keeps the process's resident set size below a user-defined
budget.}
\psentry{References}{}
\psref{1}{W.~B.~Hart, F.~Johansson, S.~Pancratz, FLINT: Fast library for
number theory, \url{https://flintlib.org}.}
\psref{2}{C.~W.~Bauer, A.~Frink, R.~Kreckel, Introduction to the GiNaC
framework for symbolic computation within the C++ programming language,
J. Symb. Comput. 33 (2002) 1--12, arXiv:cs/0004015.}
\psref{3}{J.~Beder, yaml-cpp: A YAML parser and emitter for C++,
\url{https://github.com/jbeder/yaml-cpp}.}
\psref{4}{T.~Peraro, FiniteFlow: multivariate functional reconstruction using
finite fields and dataflow graphs, JHEP 07 (2019) 031, arXiv:1905.08019.}
\psref{5}{J.~Klappert, F.~Lange, Reconstructing rational functions with
FireFly, Comput. Phys. Commun. 247 (2020) 106951, arXiv:1904.00009.}
\psref{6}{B.~Shoshany, A C++17 thread pool for high-performance scientific
computing, SoftwareX 26 (2024) 101687, arXiv:2105.00613.}
\end{programsummary}

\input{Sec1_Intro}
\input{Sec2_Prelim}
\input{Sec3_Arch}
\input{Sec3_Algorithms}
\input{Sec4_Engine}
\input{Sec5_Install}
\input{Sec6_Bench}
\input{Sec7_Summary}
\input{Sec8_Contrib}

\section*{Acknowledgements}

The authors thank X.~Chang for numerous discussions on the design and
development of the program, and X.~Huang for discussions on the benchmark
cases. The authors also thank M.~Zeng and A.~Smirnov for correspondence on the
use of FIRE and for suggestions on the manuscript. This work was supported by the National Natural Science
Foundation of China with Grants No.~12135013.

\section*{Declaration of competing interest}

The authors declare that they have no known competing financial interests or
personal relationships that could have appeared to influence the work reported
in this paper.

\section*{Data availability}

The SecRed source code and the input files used for the benchmarks are
available in the repository listed in the Program Summary. The program is also
archived in the CPC Program Library.

\bibliographystyle{elsarticle-num}
\bibliography{sec}

\end{document}

%% file: Sec1_Intro.tex
\section{Introduction}\label{sec:intro}

Precision calculations for the LHC and for rare decays rest on multi-loop
Feynman integrals. Once a process is computed beyond leading order, its
diagrams are assembled into families of scalar integrals, and every family is
reduced to a small set of master integrals with integration-by-parts (IBP)
identities~\cite{Tkachov:1981wb,Chetyrkin:1981qh,Laporta:2000dsw}. The masters
are then evaluated analytically or numerically. Their number is
finite~\cite{Smirnov:2011num}, yet choosing a basis that keeps the coefficients
small is a problem of its own~\cite{Smirnov:2020cho,Boehm:2020simple}.
Typical applications include
Higgs production at three loops~\cite{Anastasiou:2015higgs}, charged-current
Drell--Yan production at N$^3$LO~\cite{Duhr:2020dy}, two-loop five-point
amplitudes~\cite{Abreu:2019fivegluon}, and the $2\to3$ reactions on the Les
Houches precision wishlist~\cite{Huss:2026wishlist}. The reduction is the
common step, and recent reviews describe its algorithms and its
cost~\cite{Grozin:2011,Pak:2011xt,Papadopoulos:2025hoc,Smirnov:2025review,Armadillo:2026review}.

The standard Laporta algorithm~\cite{Laporta:2000dsw} turns the reduction into
the elimination of a finite system, and the size of that system sets the cost. An integral is
characterized by its \emph{depth}, the total power of the propagators in the
denominator, and its \emph{rank}, the total power of the numerator. The
algorithm chooses two bounds, $D_{\max}$ and $R_{\max}$, one layer above the
largest depth and rank among the integrals to be reduced, and applies the
identities to every integral with depth less than $D_{\max}$ and rank less
than $R_{\max}$. These integrals are the \emph{seeds}. Integrals with the same
set of propagators in the denominator form a sector. For a family with $L$
loops, $N$ propagators and $E$ independent external momenta, a sector with $k$
active propagators contains
$\binom{D_{\max}-1}{k}\binom{R_{\max}+N-k-1}{N-k}$ seeds, polynomial in the
two bounds but exponential in $k$. Every seed carries
$L(L+E)+E(E-1)/2$ identities. Each identity connects an integral to
its nearest neighbours, so the generated system is sparse, but the number of
rows and the total number of terms grow with the same combinatorial factor.
The three-loop five-point family of Ref.~\cite{Liu:2025threeloop} already lies
beyond the reach of the standard programs and required a dedicated
syzygy-based generator. At four loops the equation count reaches the
ten-billion scale: for the four-loop cut family F1 studied in Sec.~\ref{sec:bench},
a global Laporta run has to generate $1.02\times10^{10}$ equations
before the first elimination step. Holding such a system exceeds the memory
of a single node, and eliminating it at this scale takes days or months even
after it is partitioned to fit. Memory and run time therefore limit the reduction.

Three quantities set this limit: the size of the equation system
that has to be generated, the time its elimination takes, and the memory
needed to hold it. The $2\to3$ frontier is one reason these quantities
grow~\cite{Huss:2026wishlist}, since an extra external leg adds a scale and a
loop-momentum routing. Higher
powers of propagators in four-loop corrections, additional mass scales and
larger master bases have the same effect. Programs have automated the
Laporta
algorithm~\cite{Anastasiou:2004vj,Maierhoefer:2017hyi,Lee:2013mka,Lee:2014tja,Smirnov:2008iw,Smirnov:2019qkx}
and computer-algebra systems make the generation
feasible~\cite{Vermaseren:2000nd}, but in demanding calculations the reduction,
together with the reconstruction of its rational coefficients, dominates the
computing cost~\cite{Smirnov:2025review}.

Part of this cost is arithmetic, and it now has a standard treatment.
Instead of eliminating
over the rational numbers, where intermediate coefficients grow without bound,
many current programs, for example
Kira~\cite{Klappert:2020nbg,Lange:2025fba} and
FIRE~\cite{Smirnov:2019qkx,Smirnov:2025fire7}, can eliminate modulo a large
prime and recover the exact result by Chinese remaindering and rational
reconstruction. The dataflow framework of
FiniteFlow~\cite{Peraro:2019sve,Peraro:2016wsq} and the interpolation library
FireFly~\cite{Klappert:2020firefly,Klappert:2021int} supply this machinery.
This machinery is now common infrastructure rather than a distinguishing
feature of any single program. Recent work extends it: balanced reconstruction of sparse
rational functions~\cite{Belitsky:2023,Chawdhry:2024} and factorized storage of
denominators during the reduction~\cite{Smirnov:2026fuel}.

Another line of work compresses the equation system through algebraic
structure before the elimination. Unitarity-cut and syzygy methods construct
IBP relations with a
controlled propagator
degree~\cite{Gluza:2011,Schabinger:2012,Larsen:2015ped,Georgoudis:2017azurite,Bohm:2017xot,Bohm:2018hexagonbox,Kosower:2018},
the group structure exposes relations among
the identities~\cite{Lee:2008grp}, NeatIBP generates a small set from module
intersections~\cite{Wu:2023upw,Wu:2025aen}, and Blade builds a
block-triangular form and reduces spanning sector by spanning
sector~\cite{Guan:2025blade}. More recent constructions order the reduction by
a geometric relation and produce $\varepsilon$-factorized differential
equations~\cite{Bree:2026reduce}, use the reduction operators of the GKZ
system associated with a graph~\cite{Britto:2026resonance}, or fix the
coefficients from analyticity constraints with a few IBP
samples~\cite{Li:2026bootstrap}. Each of them reduces the size of the system
that reaches the elimination independently of the arithmetic used to solve it.
Their constructions, however, are expensive in their own right.

A further line of work controls the seed set. Kira~3 introduces a truncation of the seed
set, \texttt{truncate\_sp}, which keeps only the seeds expected to generate
useful equations and leaves the others as intermediate
integrals~\cite{Lange:2025fba}. Blade restricts the seed set through its block
and spanning-sector structure~\cite{Guan:2025blade}. Tube seeding restricts it
to a thin path between a target and the masters, so that the number of seeds
grows linearly rather than polynomially with the numerator
power~\cite{Berman:2026tube}. Machine-learning approaches search for the
priority function that selects the seeds to retain, by genetic programming or
language-model search~\cite{vonHippel:2025ml,Song:2026funsearch}, or by
reinforcement learning over the reduction steps~\cite{Zeng:2025rl}. Seedless
formulations avoid the seed set altogether~\cite{delaCruz:2026seedless}.

SecRed, the program presented in this article, reduces an integral family
sector by sector and, within this organization, improves seed selection,
equation-size control, elimination efficiency and memory management. The sector-wise organization is
also used by Reduze~\cite{Studerus:2009ye,vonManteuffel:2012np},
FIRE~\cite{Smirnov:2008iw,Smirnov:2019qkx,Smirnov:2025fire7} and
NeatIBP~\cite{Wu:2023upw,Wu:2025aen}. It is adopted for two
practical reasons: only the system of the sector currently being processed has to reside
in memory, and the sectors form independent tasks whose run times differ by
orders of magnitude, which makes them suitable for dynamic scheduling. Sectors run under a memory-aware scheduler that admits
tasks according to the measured resident set size and keeps the process within
a user-defined budget (Sec.~\ref{subsec:sched}). SecRed uses an adaptive seed
box with a corner cut (Sec.~\ref{subsec:atps}), can truncate the seed set with
self-correcting caps (Sec.~\ref{subsec:trunc}), applies a subsector level
filter (Sec.~\ref{subsec:silf}), and expands symmetry relations only from the
master integrals of a sector (Sec.~\ref{subsec:symmetry}). On the elimination
side, it records the elimination plan once and replays it at every prime, and
orders the equations by a head score (Sec.~\ref{sec:engine}).

The benchmarks (Sec.~\ref{sec:bench}) cover F234, a two-loop five-point
non-planar family, and F1, a four-loop four-point cut family whose global
system exceeds the memory of a single node. They test these components and
compare SecRed with other reducers. On F234 the global system remains compact,
the memory footprint modest and the wall-clock time short. On F1, SecRed
completes a reduction whose global system cannot be assembled in memory. The reductions have been cross-checked against
Kira~3~\cite{Lange:2025fba} and LiteRed~\cite{Lee:2013mka,Lee:2014tja}.
SecRed also serves as the IBP engine of AMFlow~\cite{Liu:2022chg} for the
differential-equation
method~\cite{Kotikov:1991,Tarasov:1996,Lee:2010aa,Henn:2013,Lee:2015reducingde},
and the numerical results for the target integrals agree with those obtained
with other programs. These benchmarks show that SecRed is competitive
with existing reducers and extends the range of reductions that can be
completed on a single shared-memory node.

Section~\ref{sec:prelim} recalls the definitions and identities used
throughout. Section~\ref{sec:arch} describes the architecture and the
sector-level algorithms, and Sec.~\ref{sec:engine} describes the elimination
engine. Sec.~\ref{sec:install} documents installation and usage, and
Sec.~\ref{sec:bench} reports the benchmarks. Sec.~\ref{sec:summary}
summarizes the results.

%% file: Sec2_Prelim.tex
\section{Preliminaries: integration-by-parts reduction}\label{sec:prelim}

We consider a family of Feynman integrals generated by a graph with $L$ loops,
$N$ internal lines and $E$ independent external momenta. Every integral is
labeled by a vector of integer indices $\mathbf{a}=(a_1,\dots,a_N)$,
\begin{equation}
  I(\mathbf{a}) = \int \prod_{k=1}^{L} \frac{\mathrm{d}^d l_k}{i\pi^{d/2}}\;
  \frac{1}{D_1^{a_1}\cdots D_N^{a_N}} ,
  \label{eq:integral}
\end{equation}
where $l_i$ are the loop momenta, $d$ is the space--time dimension in
dimensional regularization, and the propagators $D_k=q_k^2-m_k^2$ are
quadratic in the loop momenta, with $q_k$ a linear combination of loop and
external momenta. A negative index moves the propagator to the numerator,
$1/D_k^{a_k}=D_k^{-a_k}$ with $-a_k>0$, and $a_k=0$ means that the propagator
is absent.

\subsection{Depth, rank, level and sectors}\label{subsec:notation}

Two integers characterize the complexity of an integral,
\begin{equation}
  \Depth(\mathbf{a}) = \sum_{i:\,a_i>0} a_i , \qquad
  \Rank(\mathbf{a}) = \sum_{i:\,a_i<0} \bigl(-a_i\bigr) ,
  \label{eq:depth-rank}
\end{equation}
the \emph{depth} $\Depth$ (the total power of the propagators in the denominator) and
the \emph{rank} $\Rank$ (the total power of the numerator factors). Their sum
is the \emph{level},
\begin{equation}
  \ell(\mathbf{a}) = \Depth(\mathbf{a}) + \Rank(\mathbf{a})
  = \sum_i|a_i| .
  \label{eq:level}
\end{equation}
Integrals are classified by the \emph{sector}
\begin{equation}
  S(\mathbf{a}) = (s_1,\dots,s_N) , \qquad
  s_i = \begin{cases} 1, & a_i>0,\\ 0, & a_i\le 0 ,\end{cases}
  \label{eq:sector-array}
\end{equation}
encoded in a bitmask as the \emph{sector ID}
$\chi(S)=\sum_i s_i2^{i-1}$. We identify a sector with the set of its active
propagators, so $k\in S$ means $s_k=1$. We write
$\Pop(S)=\sum_i s_i$ for the number of active propagators (the
\emph{popcount}) and call
\begin{equation}
  \delta(\mathbf{a}) = \Depth(\mathbf{a}) - \Pop\bigl(S(\mathbf{a})\bigr)
  \label{eq:dots}
\end{equation}
the \emph{dot number} $\delta$ of the integral: the number of powers by which its
positive indices exceed one, i.e.\ the number of dots on the active
propagators. For an integral of sector $S=S(\mathbf{a})$ one has
$\ell(\mathbf{a})=\Pop(S)+\delta(\mathbf{a})+\Rank(\mathbf{a})$. A sector $S'$
is a \emph{subsector} of $S$ if
$S'\subset S$ in the bitmask sense. Sector membership depends only on the
signs of the indices: numerator factors contribute to the rank but cannot
make a propagator active.

The order in which integrals are enumerated is itself an algorithmic choice,
and SecRed uses a single key, the \emph{balance key}, in three places: the
enumeration of the seeds (the integrals to which the identities are applied),
the ordering of the subsector columns (Sec.~\ref{subsec:atps}), and the global
integral numbering. For an integral $\mathbf{a}$,
\begin{equation}
  \kappa(\mathbf{a})=\bigl(\ell(\mathbf{a}),\ \Depth(\mathbf{a}),\
  \Rank(\mathbf{a}),\ \max_i a_i,\ -\min_i a_i\bigr) ,
  \label{eq:balancekey}
\end{equation}
compared lexicographically. The level comes first so that the key is compatible
with the level cap of Sec.~\ref{subsec:atps} and the subsector level filter of
Sec.~\ref{subsec:silf}. Depth and rank then separate deep-and-narrow from
shallow-and-wide integrals, and the largest denominator power and the largest
numerator power favor one high power over several moderate ones. The
alternative \texttt{normal} order sorts by depth, then rank, with the index
vector as the final tie-break, and is used only for comparison
(Sec.~\ref{sec:f234}).

\subsection{Integration-by-parts identities and the Laporta algorithm}%
\label{subsec:ibp}

Integrals of total derivatives vanish in dimensional
regularization~\cite{Tkachov:1981wb,Chetyrkin:1981qh}. For every pair
$(l_i,q_j)$, where $l_i$ runs over the loop momenta and $q_j$ over all loop
and external momenta,
\begin{equation}
  0 = \int \prod_{k=1}^{L} \frac{\mathrm{d}^d l_k}{i\pi^{d/2}}\;
  \frac{\partial}{\partial l_i^{\mu}}
  \left( q_j^{\mu}\, \frac{1}{D_1^{a_1}\cdots D_N^{a_N}} \right) .
  \label{eq:ibp}
\end{equation}
Expanding the derivative and decomposing the quadratic factor on the
propagators, $q_j^{\mu}\partial D_k/\partial l_i^{\mu}
=\sum_n \lambda_{kn}D_n+\beta_k$, gives the index-shift form
\begin{equation}
  0 = d_{ij}\, I(\mathbf{a})
    - \sum_{k} a_k\beta_k\, I(\mathbf{a}+\mathbf{e}_k)
    - \sum_{k,n} a_k \lambda_{kn}\, I(\mathbf{a}+\mathbf{e}_k-\mathbf{e}_n) ,
  \label{eq:ibp-shift}
\end{equation}
with $d_{ij}=\partial q_j^{\mu}/\partial l_i^{\mu}$, which equals $d$ when
$q_j=l_i$ and zero otherwise. There is one identity for each of the $L(L+E)$
pairs, and each identity connects an integral only to its nearest neighbours in
index space. In addition, the invariance of the integral under Lorentz
transformations gives the Lorentz-invariance (LI)
identities~\cite{Gehrmann:1999as}. They add further relations among the
integrals of the family and can reduce the number of master integrals.
Contracting the generator of the Lorentz group with two external
momenta $p_r$ and $p_s$ yields, for
$1\le r<s\le E$,
\begin{equation}
  0 = \int \prod_{k=1}^{L} \frac{\mathrm{d}^d l_k}{i\pi^{d/2}}\;
  p_r^\mu p_s^\nu \sum_{i=1}^{L}
  \left( l_{i\mu}\frac{\partial}{\partial l_i^\nu}
       - l_{i\nu}\frac{\partial}{\partial l_i^\mu} \right)
  \frac{1}{D_1^{a_1}\cdots D_N^{a_N}} .
  \label{eq:li}
\end{equation}
The contraction is antisymmetric in $r$ and $s$, so each unordered pair
gives one independent identity and there are $E(E-1)/2$ LI identities. The
derivatives in Eq.~\eqref{eq:li} act on the propagators as in
Eq.~\eqref{eq:ibp-shift}, so the LI identities have the same shift structure.

The shift structure determines how the sector changes under an identity. In
Eq.~\eqref{eq:ibp-shift} the index $k$ is raised and $n$ is lowered, and every
term that raises $k$ carries a factor $a_k$. If $k\notin S$ then $a_k\le0$:
the terms vanish when $a_k=0$, and when $a_k<0$ the raised index $a_k+1$ is
still non-positive. Raising an index therefore never activates a propagator
outside $S$. Lowering $n$ either keeps it active ($a_n\ge2$),
removes it from the denominator ($a_n=1$, so the integral passes to the
subsector $S\setminus\{n\}$), or only increases a numerator power
($a_n\le0$). No operation activates a propagator outside $S$, and at most one
active propagator is removed at a time. The LI identities produce the same
two shift types, so the same reasoning applies to them. Consequently every
integral generated from an integral of $S$ by an IBP or LI identity belongs
to $S$ or to a strict subsector of $S$. The equations generated from the
seeds of $S$, together with the master integrals they determine, therefore
depend only on $S$ and its subsectors.

The IBP and LI identities express every integral of a family as a linear
combination of a finite set of \emph{master integrals}, with coefficients that
are rational functions of $d$ and the kinematic
invariants~\cite{Smirnov:2011num}. The Laporta
algorithm~\cite{Laporta:2000dsw} organizes the reduction as follows. The
integrals to be reduced are the \emph{targets}. Two bounds $D_{\max}$ and
$R_{\max}$ are set from the targets, one layer above the largest depth and
rank among the targets of the sector, and every integral with
$\Depth<D_{\max}$ and $\Rank<R_{\max}$ becomes a seed. Every identity is
applied to every seed, producing a linear equation. Neighbours outside
the bounds become new unknowns, while neighbours in lower sectors are moved to
the right-hand side and reduced separately. The system is ordered by a
complexity measure and solved by Gaussian elimination. The integrals that
cannot be eliminated are the masters. This is the standard construction.
Secs.~\ref{subsec:atps} and~\ref{subsec:trunc} describe how SecRed selects the
seeds and can truncate them, and Sec.~\ref{subsec:silf} how it filters the
subsector
columns. Which masters are reported depends on the targets: an irreducible
integral on which no target depends is not counted.

\subsection{Trivial sectors and symmetry relations}

The Schwinger parametrization of Eq.~\eqref{eq:integral} reads
\begin{equation}
  I(\mathbf{a}) \propto \int_0^{\infty} \prod_{i=1}^{N}
  \mathrm{d}\alpha_i\, \alpha_i^{a_i-1}\,
  \mathcal{U}(\boldsymbol{\alpha})^{d/2}\,
  \exp\left[-\frac{\mathcal{F}(\boldsymbol{\alpha})}
  {\mathcal{U}(\boldsymbol{\alpha})}\right] ,
  \label{eq:schwinger}
\end{equation}
where the first and second Symanzik
polynomials~\cite{Nakanishi:1971,Bogner:2010} are sums over spanning trees
and spanning two-forests,
\begin{align}
  \mathcal{U}(\boldsymbol{\alpha}) &= \sum_{T\in\mathcal{T}_1}\;
      \prod_{e_i\notin T}\alpha_i , \label{eq:U} \\
  \mathcal{F}(\boldsymbol{\alpha}) &= \sum_{(T_1,T_2)\in\mathcal{T}_2}
      \Bigl(\prod_{e_i\notin T_1\cup T_2}\alpha_i\Bigr)\bigl(-p_{T_1}^2\bigr)
      + \mathcal{U}(\boldsymbol{\alpha})\sum_{i} m_i^2\alpha_i ,
      \label{eq:F}
\end{align}
with $p_{T_1}$ the momentum flowing into $T_1$ through the cut lines.
$\mathcal{U}$ and $\mathcal{F}$ are homogeneous of degree $L$ and $L+1$ in the
$\alpha_i$. A sector is \emph{trivial} if all its integrals are scale-free,
i.e.\ if they carry no dimensionful scale, so that they vanish identically in
dimensional regularization. Triviality can be
decided from the Symanzik polynomials: writing $G=\mathcal{U}+\mathcal{F}$, a
sector is trivial if, after all $\alpha_i$ outside the sector are set to zero,
the linear system
\begin{equation}
  \sum_{i} b_i\,\alpha_i\,\frac{\partial G}{\partial \alpha_i} = G
  \label{eq:trivial}
\end{equation}
has a solution~\cite{Lee:2013mka,Baikov:2006}. Trivial sectors can therefore
be discarded before any equation is generated.

Sectors related by a relabeling of the propagators belong to the same integral
family. The relabeling is detected from the Symanzik polynomial
$G=\mathcal{U}+\mathcal{F}$ of Eq.~\eqref{eq:F}: setting to zero all
$\alpha_i$ outside a sector gives its \emph{restricted polynomial}, and two
sectors are symmetric if their restricted polynomials become identical under a
relabeling of the active propagators. If the relabeling is a pure
permutation that leaves the family invariant, the integral is unchanged,
\begin{equation}
  I(\mathbf{a})=I\bigl(\sigma(\mathbf{a})\bigr),
  \label{eq:sym-rel}
\end{equation}
where $\sigma$ denotes the induced permutation of the index vector. A general
relabeling is accompanied by a transformation of the loop momenta. The
relabeling is promoted to a momentum map by requiring
\begin{equation}
  q_k(l,p)=\varepsilon_k\,q'_{\sigma(k)}(l',p) ,\qquad
  l'=Ml+Kp ,\qquad \varepsilon_k=\pm1 ,
  \label{eq:sym-flow}
\end{equation}
where $q_k$ is the momentum of propagator $k$, $\sigma$ is the relabeling, and
$M$ and $K$ are constant matrices acting on the loop and external momenta.
Rewriting the propagators in the transformed basis gives a linear relation
among integrals with shifted indices,
\begin{equation}
  \sum_{\mathbf{n}\in\mathcal{N}(\mathbf{a})}
  c_{\mathbf{n}}(\mathbf{a})\, I(\mathbf{a}+\mathbf{n})=0,
  \label{eq:sym-rel-general}
\end{equation}
where $\mathcal{N}(\mathbf{a})\subset\mathbb Z^N$ is a finite set of shift
vectors and the coefficients $c_{\mathbf{n}}(\mathbf{a})$ are rational
functions of $d$ and the kinematic invariants. For a pure permutation with
$\sigma(\mathbf{a})\ne\mathbf{a}$,
$\mathcal{N}(\mathbf{a})$ contains at most the two shifts $\mathbf{n}=0$ and
$\mathbf{n}=\sigma(\mathbf{a})-\mathbf{a}$, and
Eq.~\eqref{eq:sym-rel-general} reduces to Eq.~\eqref{eq:sym-rel}.
Equations~\eqref{eq:sym-rel} and~\eqref{eq:sym-rel-general} are the
\emph{symmetry relations}~\cite{Studerus:2009ye,Wu:2025symmetry}. A symmetry
relation that contains a master $I_M$ can be solved for $I_M$,
\begin{equation}
  I_M=\sum_{i\ne M} c_i\, I_i ,
  \label{eq:sym-remove}
\end{equation}
where the sum runs over the other integrals in the relation, so $I_M$ is not
an independent master. When the relabeling connects different
sectors, the relation links integrals across the sector boundary. The
relations are added to the equation system before the elimination.
Sec.~\ref{subsec:symmetry} describes how SecRed constructs them.

\subsection{Finite-field methods and reconstruction}

Elimination over the rational numbers, or with symbolic polynomials, suffers
from intermediate expression swell. Finite-field
methods~\cite{Peraro:2016wsq,vonManteuffel:2014ixa} remove this limitation by
carrying out the elimination modulo a large prime $p$, repeating it for
several primes, and reconstructing the exact rational function from the
modular images~\cite{Kant:2013vta}. Given residues $r_i\equiv a\pmod{p_i}$
for pairwise coprime moduli, the Chinese remainder theorem gives
\begin{equation}
  \begin{aligned}
    a &\equiv \sum_{i} r_i\, M_i\, (M_i^{-1}\bmod p_i) \pmod{P} ,\\
    M_i &= P/p_i ,\qquad P=\prod_i p_i .
  \end{aligned}
  \label{eq:crt}
\end{equation}
A rational number $a=u/v$ with $\gcd(u,v)=1$ and $|u|,|v|<\sqrt{P/2}$ is
uniquely recovered from its residue modulo $P$ by the extended Euclidean
algorithm~\cite{Monagan:2004}. Polynomial dependence is recovered by Newton
interpolation in one variable, by the Zippel
algorithm~\cite{Zippel:1979,Zippel:1990} in several variables, or by balanced
reconstruction of numerator and denominator~\cite{Klappert:2021int,Belitsky:2023,Chawdhry:2024}.
FireFly and FiniteFlow provide these algorithms, and SecRed uses them for the
reconstruction in the global stage (Sec.~\ref{sec:arch}). The number of primes
is determined by the size of the largest numerator and denominator and ranges
from a few dozen to a few hundred.

%% file: Sec3_Arch.tex
\section{Architecture and algorithms for sector-wise reduction}\label{sec:arch}

SecRed reduces an integral family in two stages. The first stage processes the
non-trivial sectors independently and in parallel, determines their local
master integrals, and retains only the linearly independent equations. The
second stage assembles those equations into one global system, eliminates it,
and reconstructs the reduction coefficients. This section describes the
architecture and the four sector-level algorithms. The elimination engine is
the subject of Sec.~\ref{sec:engine}.

\subsection{Overall architecture and data flow}\label{subsec:overview}

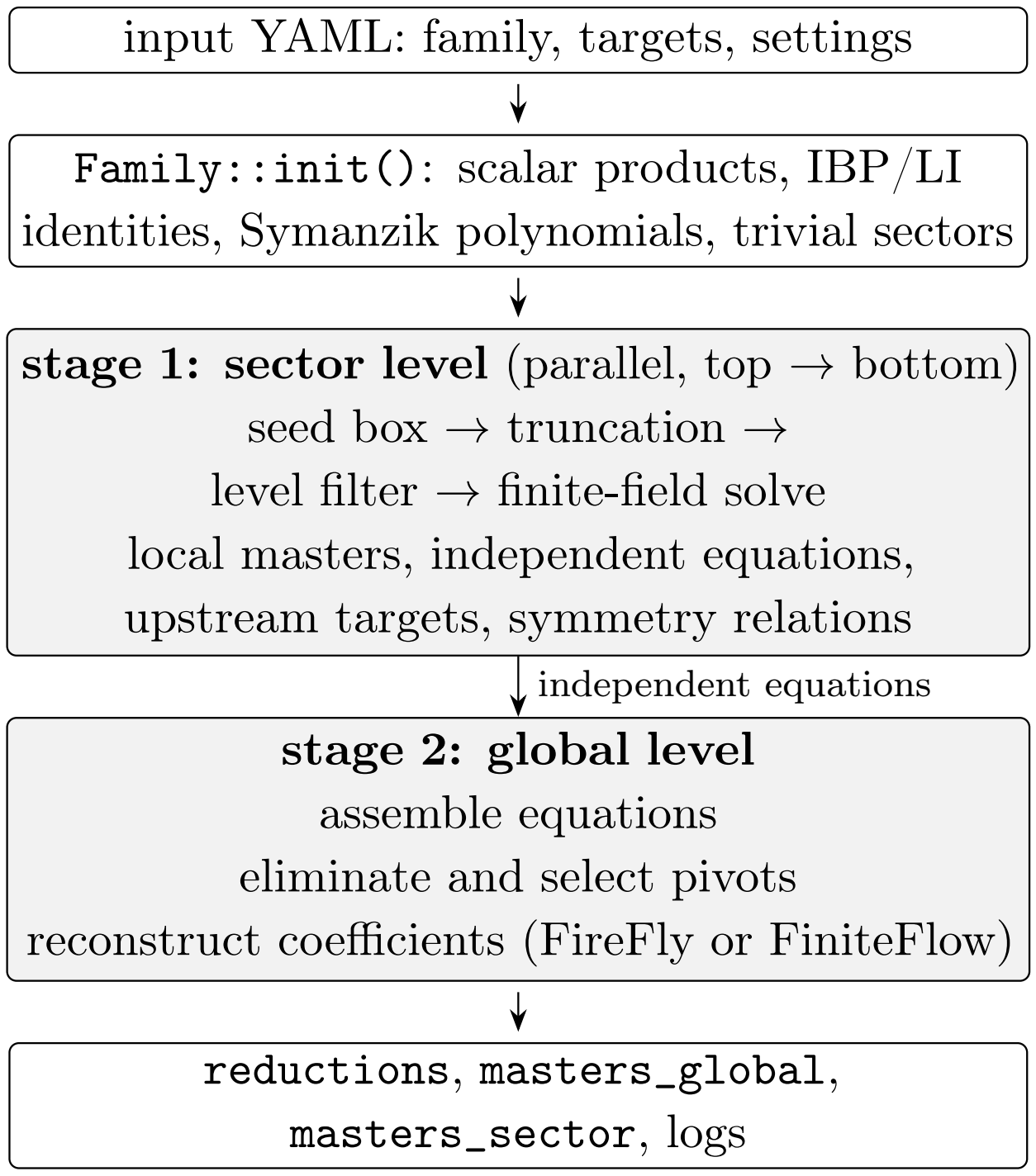
\begin{figure}[htbp]
\centering
\resizebox{8.2cm}{!}{%
\begin{tikzpicture}[
  node distance=0.45cm,
  box/.style={draw, rounded corners=2pt, align=center, inner sep=2.5pt,
              font=\footnotesize\linespread{1.12}\selectfont, text width=7.4cm},
  stage/.style={draw, rounded corners=2pt, fill=gray!10, align=center,
              inner sep=3pt, font=\footnotesize\linespread{1.12}\selectfont,
              text width=7.4cm},
  arr/.style={-Stealth, shorten <=2.2pt, shorten >=2.2pt}
]
\node[box] (yaml) {input YAML: family, targets, settings};
\node[box, below=of yaml] (init) {\texttt{Family::init()}: scalar products, IBP/LI
  identities, Symanzik polynomials, trivial sectors};
\node[stage, below=of init] (s1)
  {\textbf{stage 1: sector level} (parallel, top $\to$ bottom)\\
   seed box $\to$ truncation $\to$ level filter $\to$ finite-field solve\\
   local masters, independent equations, upstream targets, symmetry relations};
\node[stage, below=of s1] (s2)
  {\textbf{stage 2: global level}\\
   assemble equations\\
   eliminate and select pivots\\
   reconstruct coefficients (FireFly or FiniteFlow)};
\node[box, below=of s2] (out) {\texttt{reductions}, \texttt{masters\_global},
  \texttt{masters\_sector}, logs};
\draw[arr] (yaml) -- (init);
\draw[arr] (init) -- (s1);
\draw[arr, shorten <=0pt, shorten >=0pt]
  (s1) -- node[right, font=\scriptsize] {independent equations} (s2);
\draw[arr] (s2) -- (out);
\end{tikzpicture}}
\caption{Two-stage data flow of SecRed. Stage~1 is local to each sector and
parallel, and stage~2 is the only step that sees the whole family.}
\label{fig:dataflow}
\end{figure}

Figure~\ref{fig:dataflow} shows the data flow. Stage~1 processes the
non-trivial sectors in parallel, ordered by decreasing popcount. Each sector is reduced
locally, and the subsector integrals that its retained equations involve are
forwarded downwards as targets.

Stage~1 does not compute reduction coefficients. It determines which equations
are independent, which integrals are masters of a sector, and which subsector
integrals a higher sector needs. The arithmetic is
therefore finite-field arithmetic only, and the symbolic coefficients are
regenerated for the retained equations before they are written to the global
container.

As shown in Sec.~\ref{subsec:ibp}, the equations generated from the seeds of a
sector $S$ involve only $S$ and its subsectors. Ordering the unknowns by
popcount, then by depth and rank, makes the global matrix block triangular,
with each block referring only to itself and to blocks below it. This permits
the top-down traversal: a sector can be reduced as soon as its parents
have finished, and sectors of the same layer that are not in a subsector
relation are independent.

The information that travels downwards is the list of \emph{upstream targets}.
When a sector $S_{\mathrm{p}}$ finishes, every subsector integral that appears
in its retained equations and is not reducible inside $S_{\mathrm{p}}$ becomes a
target of the sector to which it belongs. The receiving sector then treats it
like a target supplied by the user: its reduction must be completed there. The
shift argument of Sec.~\ref{subsec:ibp} shows that such an integral has depth
at most the largest depth among the targets of $S_{\mathrm{p}}$: reaching a
proper subsector lowers one active denominator, which offsets the at most one
unit that the raised index can add.
In practice it lies one level below the box top of $S_{\mathrm{p}}$. The initial
depth bound of the child therefore does not exceed that of the parent. The propagation is
exact: every integral that the global stage may need is either reduced in the
sector where it was generated or forwarded to a sector that can reduce it.

A sector-level reduction is nevertheless not complete in the global sense.
The equations of $S$ may connect integrals of several different subsectors:
after the integrals of $S$ itself have been eliminated, what remains can be a
relation among integrals of two or more subsectors. Such a relation cannot be
obtained by reducing either subsector alone, because the equations of a
subsector never contain integrals of a sibling. Relations of this type are
called \emph{magic relations}~\cite{Klappert:2020nbg}. They become available
only in stage~2, when the equations retained from all sectors are assembled
and the subsector integrals become common unknowns. The sector-level master
set is therefore a superset of the final one, and the global stage removes the
redundancy. This division allows the sector stage to remain local and
inexpensive without sacrificing completeness.

Within a sector, the algorithms of Secs.~\ref{subsec:atps}--\ref{subsec:symmetry}
form a pipeline. Each stage can only shrink the set of objects passed to the
next:
\begin{equation}
  \mathcal{B}(S;D_{\max},R_{\max})
  \;\xrightarrow{\text{trunc.}}\;
  \mathcal{B}_{\mathrm{true}}(S)
  \;\xrightarrow{\text{filter}}\;
  \mathcal{C}(S)
  \;\xrightarrow{\text{elim.}}\;
  \mathcal{E}(S) .
  \label{eq:pipeline}
\end{equation}
Here $\mathcal{B}$ is the seed box, $\mathcal{B}_{\mathrm{true}}$ the true
seeds, $\mathcal{C}$ the columns that enter the elimination, and
$\mathcal{E}$ the independent equations that leave it.
The seed box is the conservative starting point, determined from the requests
of the sector. The optional truncation removes seeds that are expected to be
intermediate integrals, without removing their columns. The level filter
removes subsector columns far below the box top. The elimination then acts on
the remaining system and returns the local masters and the independent
equations. Symmetry relations are added after the local masters are known,
since they are expanded from those masters only, and any integral they involve
in another sector is forwarded like an upstream target. Each arrow in
Eq.~\eqref{eq:pipeline} is a place where the system can be made smaller
without changing the final answer.

Each sector is solved once to determine its linear structure. The coefficients
are evaluated at a single point modulo a \emph{learn prime} $p_0$, the first
of a fixed list of primes just below $2^{31}$, and the solve returns three
results: the local masters, the independent equations, and the dependencies
between them from forward selection~\cite{Lange:2025fba}. All three are
properties of the sparsity pattern rather than of the coefficient values. A
coefficient that is nonzero generically vanishes at the learn point with
probability at most $D/p_0$, where $D$ is the degree of its numerator, and a
union bound over the coefficients of the local system bounds the probability
that a structural decision differs from the generic one. The decision is
checked as well. Forward selection re-solves the selected equations at $p_0$
and requires every target to reduce to masters, repairing a missing pivot or
falling back to the full closure when it cannot, and a stored coefficient that
vanishes at a later prime sends the replay to a fresh elimination order
(Sec.~\ref{subsec:plan}). A wrong structural decision therefore triggers a
fallback or a rebuild rather than propagating unchecked. The plan that records
the decisions is reused at every prime.

The coefficients of the selected equations are regenerated afterwards. Only
the equations that survived forward selection are regenerated, not the raw
system. Writing $E_{\mathrm{fwd}}(S)$ for the number of equations forwarded by
sector $S$ and $R_{\mathrm{gen}}(S)$ for the number generated, the ratio is
\begin{equation}
  \frac{\sum_S E_{\mathrm{fwd}}(S)}{\sum_S R_{\mathrm{gen}}(S)}
  \;\approx\;
  \frac{\text{equations to global}}{\text{raw equations}} ,
  \label{eq:regen-ratio}
\end{equation}
The relation is approximate because the global count also contains the
symmetry relations, which are expanded from the local masters rather than
generated from the seeds. On F234 they are $204$ of the $1{,}162{,}358$
forwarded equations, so the difference is negligible. The ratio
is $8.0\%$ on F234 ($1{,}162{,}358$ of $14{,}478{,}997$) and $3.1\%$ on F1 at
$n=0$ ($34{,}557{,}830$ of $1.11\times10^{9}$), with both families defined in
Sec.~\ref{sec:bench}. More than $90\%$ of the generated equations never
receive a symbolic coefficient.

For a symbolic system the regeneration uses a row table shared by the whole
family. For every IBP identity it stores the coefficient of each index shift
as a polynomial in the seed indices, so the coefficient of a particular row is
obtained by evaluating that polynomial at the row's index vector rather than
by expanding the identity again. Identities whose coefficients are not
polynomial in this form fall back to GiNaC and are counted separately. For a
numeric system the coefficients are not regenerated at all: the selected rows
are replayed at each prime by the elimination plan. In both cases the
structural work of stage~1 is done once, and only the retained equations carry
coefficient data into stage~2.

%% file: Sec3_Algorithms.tex
\subsection{Adaptive seed box and corner cutting}\label{subsec:atps}

For a sector $S$ with bounds $D_{\max}$ and $R_{\max}$, the system has one
column per integral with $\Depth(I)\le D_{\max}$ and
$\Rank(I)\le R_{\max}$. An integral generates equations only when
\begin{equation}
  \Depth(I)<D_{\max},\qquad \Rank(I)<R_{\max},\qquad
  \ell(I)<\ell_{\max} ,
  \label{eq:seedbox-admit}
\end{equation}
where $\ell_{\max}$ is a cap on the level. The first two conditions define a
rectangle in the depth--rank plane, and the third cuts off its upper-right
corner.
An integral on the far face, with $\Depth=D_{\max}$ or $\Rank=R_{\max}$, is
kept as a column because it can appear as a term in an equation, but it seeds
no equation of its own.
The starting box is determined by the requests of the sector,
$\mathcal{Q}(S)$ for user targets and $\mathcal{P}(S)$ for upstream targets,
\begin{equation}
  \begin{aligned}
    D_{\max}&=\max_{\mathbf{t}\in\mathcal{Q}(S)\cup\mathcal{P}(S)}
    \Depth(\mathbf{t})+1 ,\\
    R_{\max}&=\max\Bigl(1,\ \max_{\mathbf{t}\in\mathcal{Q}(S)\cup\mathcal{P}(S)}
    \Rank(\mathbf{t})+1\Bigr) ,
  \end{aligned}
  \label{eq:box-bounds}
\end{equation}
and the level cap is placed one layer below the corner,
\begin{equation}
  \ell_{\max}=D_{\max}+R_{\max}-2 .
  \label{eq:levelcap}
\end{equation}
Two request sets enter Eq.~\eqref{eq:box-bounds}: the user targets
$\mathcal{Q}(S)$ and the upstream targets $\mathcal{P}(S)$ registered by the
parents. A box built from $\mathcal{Q}(S)$ alone is the \emph{target box}, and
the union gives the \emph{request box}, which is the one used.
The largest level admitted by the box is the \emph{box top}
\begin{equation}
  \ell_{\mathrm{top}}
  =\min\bigl(D_{\max}+R_{\max}-2,\ \ell_{\max}-1\bigr) .
  \label{eq:boxtop}
\end{equation}
Initially, Eq.~\eqref{eq:levelcap} gives
$\ell_{\mathrm{top}}=D_{\max}+R_{\max}-3$. The
corner integral, at depth $D_{\max}-1$ and rank $R_{\max}-1$, is outside the
box.

The corner cut is exact and inexpensive to evaluate. For a sector with
$t=\Pop(S)$ active and $m=N-t$ inactive propagators, the number of seeds of
depth $\Depth$ and rank $\Rank$ is
$\binom{\Depth-1}{t-1}\binom{\Rank+m-1}{m-1}$. Summing over the rectangle gives
\begin{equation}
  \begin{aligned}
    N_{\mathrm{box}}&=\binom{D_{\max}-1}{t}\binom{R_{\max}+m-1}{m} ,\\
    N_{\mathrm{cut}}&=\binom{D_{\max}-2}{t-1}\binom{R_{\max}+m-2}{m-1} ,
  \end{aligned}
  \label{eq:seedcounts}
\end{equation}
so that the cut removes the fraction
\begin{equation}
  f_{\mathrm{cut}}=\frac{N_{\mathrm{cut}}}{N_{\mathrm{box}}}
  =\frac{t}{D_{\max}-1}\,\frac{m}{R_{\max}+m-1} .
  \label{eq:fcut}
\end{equation}
For a representative F234 sector with $t=5$, $D_{\max}=9$, $R_{\max}=4$ and
$m=4$, this gives $f_{\mathrm{cut}}=20/56\approx35.7\%$: more than a third of
the seeds are not generated, without changing which integrals are reachable.

The initial box is a starting point, not a guarantee. In a round the sector
system for the current box is built and solved, and the local masters that the
requests depend on are then inspected. Because the elimination orders integrals
by the balance key of Eq.~\eqref{eq:balancekey}, whose first criterion is the
level, a master needed by a target has no larger level than the target. The
corner cut therefore has three cases. If the corner holds no target, no needed
master can appear there, so the cut is safe and the first round settles. If the
corner holds a target but no master appears there, the target is expressed
through integrals below the corner, so the cut is safe and the first round
settles. If a master appears at the corner, the corner integral has to generate
an equation. Only this third case produces a master above the top,
\begin{equation}
  \mathrm{exceeded}(m)=\bigl[\,\ell(m)>\ell_{\mathrm{top}}\,\bigr] ,
  \label{eq:exceeded}
\end{equation}
so the round is discarded and $\ell_{\max}$ is raised by one level. The raised
cap admits the corner, and the second round settles. Whichever round settles,
its system is the one retained. The discarded first round is cheaper because
its box is smaller. Algorithm~\ref{alg:atps} states the procedure.

\begin{algorithm}[htbp]
\caption{Adaptive seed box (one sector $S$)}\label{alg:atps}
\begin{algorithmic}[1]
\Require $S$, requests $\mathcal{Q}(S)\cup\mathcal{P}(S)$
\State $D_{\max}\gets\max_{\mathbf t}\Depth(\mathbf t)+1$;\quad
       $R_{\max}\gets\max(1,\max_{\mathbf t}\Rank(\mathbf t)+1)$
\State $\ell_{\max}\gets D_{\max}+R_{\max}-2$;\quad
       $\ell_{\mathrm{top}}\gets\min(D_{\max}+R_{\max}-2,\ell_{\max}-1)$
\For{round $=1,2$}
  \State enumerate the columns, select the seeds of Eq.~\eqref{eq:seedbox-admit}
  \State apply truncation (Sec.~\ref{subsec:trunc}) and level filter
         (Sec.~\ref{subsec:silf}), and solve the sector system
  \If{no needed master $m$ has $\ell(m)>\ell_{\mathrm{top}}$}
    \State \textbf{settle}: return local masters and independent equations
  \EndIf
  \State raise $\ell_{\max}$ by one level to admit the corner
  \State discard the round
\EndFor
\end{algorithmic}
\end{algorithm}

The measurements on F234 at $n=1$ with the truncation off
(Sec.~\ref{subsec:trunc}) show how the corner cut performs. It succeeds at the
first round in $208$ of the $219$ non-trivial sectors, a rate of $95.0\%$. The
remaining $11$ need one extension. Across the family, the cut
removes $24.5\%$ of the seeds before truncation, from $1{,}421{,}510$ to
$1{,}073{,}519$.

\subsection{Adaptive seed truncation}\label{subsec:trunc}

The adaptive seed box of Sec.~\ref{subsec:atps} follows a conservative rule: every
unreduced subsector integral that appears in a selected independent equation is
registered as an upstream target and must be expressed in its own sector. The
rule leaves no intermediate integral unreduced, but the sector stage has no
global view. An intermediate integral that a sector is required to reduce need
not lie on a reduction path that the global stage uses, and the box
covers requests that the sector stage may not need.

The practice in Kira~3~\cite{Lange:2025fba} is more aggressive. Its seed
truncation (\texttt{truncate\_sp}) generates equations only from seeds inside
depth and rank caps, and keeps the integrals outside them as intermediate
integrals that remain columns without generating equations. The caps can be
tightened so that only the user targets are guaranteed to reduce. The upstream
targets are then ignored for the moment and forwarded to the global stage as
irreducible integrals. The expectation is not that the global stage reduces
these integrals, but that it reduces the user targets along paths that avoid
them. The risk is that a truncation that is too tight removes a reduction path
that the global stage cannot reconstruct, and Kira~3 relies on manual tuning of
the truncation to reach a minimal master basis.

The seed box fixes which integrals must be present in the system, not which of
them must generate equations. The truncation makes that second choice,
separating the \emph{true seeds}, which generate equations, from the
\emph{intermediate integrals}, which remain columns without equations. SecRed
implements the truncation as a switch, off by default, and makes the choice
adaptive rather than manual when it is on. Within a sector, a refused seed
that the round needs is detected from the returned masters, and the cap on the
corresponding axis is lifted for the next round. Each sector also records the
requests that its truncation refuses. If a final master in the settled global
basis is one of these requests, it can be traced back to the sector whose cut
was too tight, and the truncation rule is relaxed before the family is reduced
again. The caps are computed from the input targets as follows.

Let $\mathcal{Q}_{\mathrm{in}}$ be the targets read from the input file, and
let $t_{\mathrm{top}}$ be the number of propagators of the top sector, whose
active propagators are exactly those that carry a positive power in at least
one input target. Define
\begin{equation}
  \begin{aligned}
    \rho_{\max}&=\max_{\mathbf{t}\in\mathcal{Q}_{\mathrm{in}}}\Rank(\mathbf{t}) ,
    \qquad
    D_\delta=\max_{\mathbf{t}\in\mathcal{Q}_{\mathrm{in}}}\delta(\mathbf{t}) ,\\
    \rho_{\mathrm{exc}}(S)&=\max_{\substack{\mathbf{t}\in\mathcal{Q}_{\mathrm{in}}\\
    S(\mathbf{t})=S}}\Rank(\mathbf{t}) ,
  \end{aligned}
  \label{eq:trunc-derivation}
\end{equation}
with $\rho_{\mathrm{exc}}(S)=0$ if no input target belongs to $S$. The level
offset and the rank cap of the sector are
\begin{equation}
  \begin{aligned}
    \ell_{\mathrm{off}}&=\max\bigl(0,\ t_{\mathrm{top}}-\rho_{\max}-o\bigr) ,\\
    \rho_{\mathrm{keep}}(S)&=\max\bigl(1,\ t_S-\ell_{\mathrm{off}}+1,\ \rho_{\mathrm{exc}}(S)\bigr) ,
  \end{aligned}
  \label{eq:trunc-keep}
\end{equation}
where $t_S=\Pop(S)$ and $o$ is the offset \texttt{seed\_level\_offset}. The
offset is $0$ if all input targets have $D_\delta=0$ and $1$ otherwise, unless
the input sets it explicitly. The two truncation caps are
\begin{equation}
  \begin{aligned}
    D_{\mathrm{cut}}&=\min\bigl(D_{\max},\ t_S+D_\delta+1\bigr) ,\\
    R_{\mathrm{cut}}&=\min\bigl(R_{\max},\ \rho_{\mathrm{keep}}(S)+1\bigr) ,
  \end{aligned}
  \label{eq:trunc-caps}
\end{equation}
and a seed generates an equation only if, in addition to
Eq.~\eqref{eq:seedbox-admit},
\begin{equation}
  \Depth(I)<D_{\mathrm{cut}} ,\qquad \Rank(I)<R_{\mathrm{cut}} .
  \label{eq:trunc-admit}
\end{equation}
A truncated seed is not removed from the system. It remains a column and
receives a column weight, so a target can still be expressed through it. It
does not generate equations. The truncation is therefore
self-correcting. If a truncated intermediate integral proves to be needed, it
appears among the masters returned by the round, and the program can detect
that the truncation refused a needed seed and lift the cap on the
corresponding axis for the next round.

Which refused masters trigger a correction is controlled by
\texttt{probe\_mode}, which is consulted only while the truncation is on. In
targets mode, only masters on the
reduction path of an input target are inspected. A master needed only by an
upstream target is accepted as a local master and left to the global stage.
In all mode, every refused master triggers a correction. Targets mode
keeps the truncation adaptive rather than recursive: an upstream
target cannot force a sector to enlarge its box, and then its parents to do
the same, along the whole request chain. The price is that the local basis may
contain a few integrals that a larger box would have eliminated. The global
stage can still remove them if the equations suffice.

Two feedback checks act after the global basis has settled. Let
\begin{equation}
  D_\delta^{\mathrm{settled}}=\max_{\mathbf{m}\in\mathcal{M}}\delta(\mathbf{m})
  \label{eq:settled-dots}
\end{equation}
be the largest dot number in the final basis $\mathcal{M}$. If
$D_\delta^{\mathrm{settled}}>D_\delta$, the truncation has removed a dot layer
that the basis needs. $D_\delta$ is then raised to the observed value and the whole
run is repeated. If the settled basis contains a request that the truncation
refused, the request truncation is lifted as a whole and the run is repeated.
Lifting single requests does not converge, because releasing one request
changes the targets propagated to its subsectors. Both checks share a limit
of four passes, and the final pass is the reported result. Both feedback
reruns occur in the samples distributed with the program: \texttt{tennis}
raises the dots limit and \texttt{L04} lifts the request truncation.

On F234, the input targets give
$t_{\mathrm{top}}=8$, $\rho_{\max}=4$ and $D_\delta=2$, so the automatic offset
$o=1$ gives $\ell_{\mathrm{off}}=8-4-1=3$. For a five-line subsector that carries no input
target, $\rho_{\mathrm{exc}}=0$ and Eq.~\eqref{eq:trunc-caps} gives
\begin{equation}
  \begin{aligned}
    D_{\mathrm{cut}}&=\min(D_{\max},\,5+2+1)=8 ,\\
    R_{\mathrm{cut}}&=\min\bigl(R_{\max},\,\max(1,5-3+1)+1\bigr)=4 ,
  \end{aligned}
  \label{eq:trunc-example}
\end{equation}
so the true seeds of that sector are the box seeds with $\Depth\le7$ and
$\Rank\le3$, while the rank-4 seeds remain columns without
generating equations. A sector that does carry a rank-4 input target has
$\rho_{\mathrm{exc}}=4$, and the rank cap is lifted back to $R_{\max}$. This is
the exception mechanism in its simplest form. With the truncation switched
off, the run admits the whole seed box and generates $14{,}478{,}997$ raw
equations. With the switch on in targets mode, $182$ of the $219$ non-trivial
sectors are truncated, no cap is lifted by the truncation probe, and the run generates
$1{,}113{,}377$ raw equations, $92.3\%$ fewer. Both runs settle on the same
$309$ masters.

When the truncation is on, it also masks the corner cut. In a truncated sector at least one
truncation cap lies below the corresponding box bound, so every seed inside the
caps has $\ell\le D_{\max}+R_{\max}-3$, at or below the box top. The corner cut
removes only seeds above that top, so it changes nothing in the generated system
of the $182$ truncated sectors. Its effect is confined to the $37$ sectors
without truncation. The cut still contributes: on F234 with the truncation on,
it removes $9.6\%$ of the seeds, from $94{,}781$ to $85{,}681$, and acts in
$36$ of those $37$ sectors.

\subsection{Subsector level filtering}\label{subsec:silf}

The subsector filter rests on a linear-algebra fact. In Gaussian elimination performed
from left to right, the pivot of a column depends only on that column and the
columns before it, so deleting columns to the right cannot change which of the
earlier columns become masters.

The filter is motivated by the cost of the local elimination. The local system
is ordered so that the sector's own integrals come first and the subsector
columns after them. Most of the columns are subsector integrals generated by
the identities: a single equation can refer to several different subsectors,
and in a sector with many active propagators they can exceed $80\%$ of the
columns.
Eliminating the complete local system of one sector can be even more demanding
than eliminating the whole global system, because the subsector integrals at
the back of the equations bring a large number of terms and much complexity
into the local elimination, while they do not affect which of the sector's own
integrals at the front become masters. A natural approach is to filter all
subsector integrals. This removes the subsector complexity without changing
the mathematical structure of the sector's own reduction (FIRE, for
example~\cite{Smirnov:2025fire7}). It also removes information, however. If
all subsector integrals are masked, the magic relations of
Sec.~\ref{subsec:overview}, which appear when the sector's own integrals
cancel among themselves, collapse to zero, and the global stage loses the
information it uses to remove redundant masters. The problem is therefore a
balance between efficiency and information: the filter should remove enough
subsector integrals to make the local elimination faster while losing as few
potential magic relations as possible.

Filtering a column means omitting its term from every equation and keeping the
rest, i.e.\ projecting the system onto the retained columns. We work in the
balance ordering of Sec.~\ref{subsec:notation} and keep only the subsector
integrals in the top $n$ levels, counted by their distance to the box top.

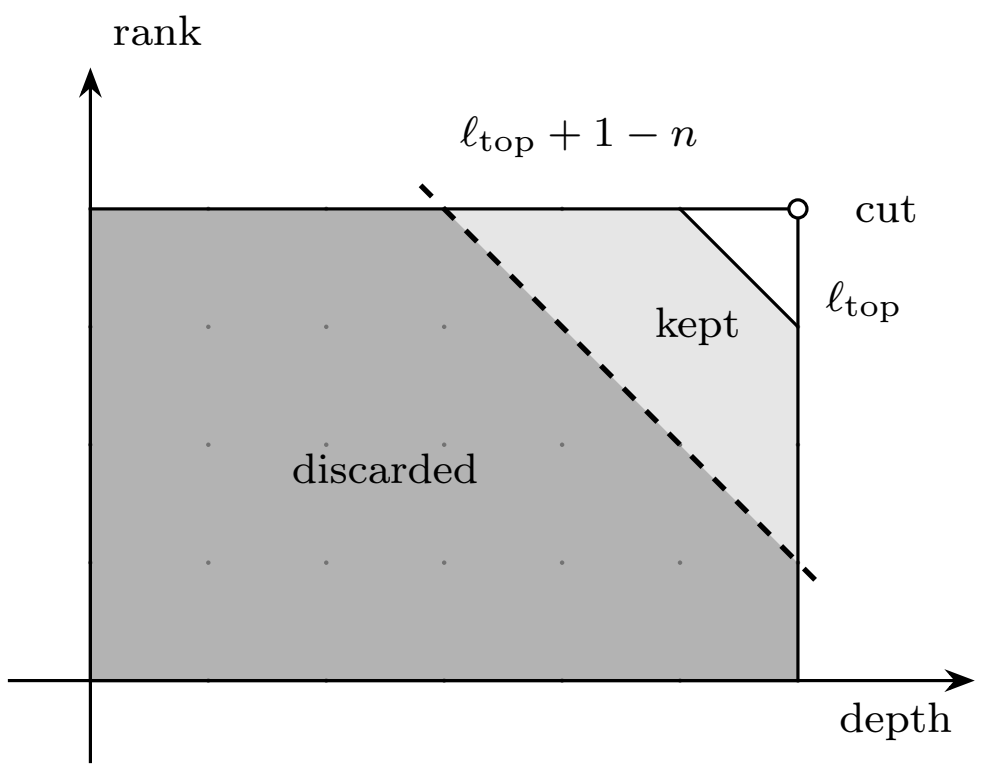
\begin{figure}[htbp]
\centering
\begin{tikzpicture}[
    x=0.78cm, y=0.78cm,
    every node/.style={font=\scriptsize},
    axis/.style={-Stealth, line width=0.6pt},
    kept/.style={fill=black!10},
    dropped/.style={fill=black!30},
    cutl/.style={line width=1pt, dashed},
  ]
  \fill[dropped] (0,0) -- (6,0) -- (6,1) -- (3,4) -- (0,4) -- cycle;
  \fill[kept] (6,1) -- (3,4) -- (5,4) -- (6,3) -- cycle;
  \foreach \x in {0,...,6}{%
    \foreach \y in {0,...,4}{%
      \fill[black!55] (\x,\y) circle (0.42pt);}}
  \draw[axis] (-0.7,0) -- (7.5,0) node[anchor=north east] {depth};
  \draw[axis] (0,-0.7) -- (0,5.2) node[anchor=south west] {rank};
  \draw[line width=0.6pt] (0,0) rectangle (6,4);
  \draw[line width=0.6pt] (5,4) -- (6,3);
  \node[anchor=south west] at (6.05,2.85) {$\ell_{\mathrm{top}}$};
  \draw[fill=white, line width=0.6pt] (6,4) circle (1.7pt);
  \node[anchor=west] at (6.3,4) {cut};
  \draw[cutl] (2.8,4.2) -- (6.2,0.8);
  \node[anchor=south west] at (2.95,4.25) {$\ell_{\mathrm{top}}+1-n$};
  \node at (2.5,1.8) {discarded};
  \node at (5.15,3.0) {kept};
  \useasboundingbox (-0.8,-0.95) rectangle (7.6,5.4);
\end{tikzpicture}
\caption{Depth--rank plane of a sector. The rectangle is the seed box of
Eq.~\eqref{eq:seedbox-admit}. The level cap removes the corner integral at
depth $D_{\max}-1$ and rank $R_{\max}-1$ (open circle). It lies above the box
top $\ell_{\mathrm{top}}$. The subsector filter retains the band
$\ell\ge\ell_{\mathrm{top}}+1-n$ (light) and discards the lower band (dark).
The sector's own columns are kept everywhere inside the box.}
\label{fig:boxgrid}
\end{figure}

A subsector integral $J$ is retained when its level satisfies
\begin{equation}
  \ell(J)=\Depth(J)+\Rank(J)\ \ge\ \ell_{\mathrm{top}}+1-n ,
  \label{eq:silf}
\end{equation}
and discarded otherwise. The offset $n$ is \texttt{subsector\_filter\_n},
default $1$. At $n=0$ the threshold is $\ell_{\mathrm{top}}+1$, so every
subsector column is discarded and the system contains only the sector's own
integrals. At $n=1$ only the top level is kept. As $n$ grows, the retained band widens until
the unfiltered Laporta system is recovered. We denote this limit by
$n=\infty$, although the value is finite in practice. Figure~\ref{fig:boxgrid}
shows the retained band and the corner cut in the depth--rank plane.

The effect is measured by scanning $n$ on F234 (Table~\ref{tab:nscan}). At
$n=0$ the basis has $314$ masters. At $n=1$ it drops to $309$ and stays there
for $n=2,3,\infty$, while the wall-clock time grows from $1$m$11$s at $n=0$ to
$12$m$28$s in the unfiltered limit. The five extra masters at $n=0$ are the price of discarding all
subsector columns: they are reducible integrals that the cross-subsector
relations of Sec.~\ref{subsec:overview} would have removed. The default $n=1$
is the smallest offset that closes the reduction on this family. On F1 the
trade-off reverses: $n=0$ gives $566$ masters in $1$h$14$m$43$s, while $n=1$ gives
$559$ masters but takes $3$h$24$m$30$s, so the larger family is run at $n=0$
(Sec.~\ref{sec:f1}).

\subsection{Symmetry relations from local masters}\label{subsec:symmetry}

Besides the magic relations of Sec.~\ref{subsec:overview}, symmetry relations
form a second class of supplementary relations that remove redundant masters.
In the standard construction one relation is generated from every seed, so the
number of relations is proportional to the size of the seed set.

A relation is obtained by rewriting each propagator of the seed in terms of
the propagators $D'_j$ of the target sector. The map of Eq.~\eqref{eq:sym-flow}
gives the loop-momentum transformation $l=M^{-1}(l'-Kp)$ and the
correspondence $k\leftrightarrow\sigma(k)$ between the propagators.
Substituting the transformation into $D_k=q_k^2-m_k^2$ expresses the
propagator as
\begin{equation}
  D_k=\sum_j f(k,j)\,D'_j+c_k ,
  \label{eq:sym-prescription}
\end{equation}
where $f(k,j)$ is the coefficient of $D'_j$ and $c_k$ is the constant
remainder, which contains the mass shift of the mapped propagator. A seed
that carries numerator powers produces a relation whose number of terms grows
with the rank of the seed. If propagator $k$ is replaced by $f_k$ terms,
where $f_k$ is the number of nonzero coefficients $f(k,j)$ of
Eq.~\eqref{eq:sym-prescription}, a numerator power $|a_k|$ multiplies the
number of terms of the relation by
$\binom{|a_k|+f_k-1}{f_k-1}$, so
\begin{equation}
  N_{\mathrm{terms}}(I)\ \le\ \prod_{k:\,a_k<0}
  \binom{|a_k|+f_k-1}{f_k-1} ,
  \label{eq:sym-cost}
\end{equation}
which is smallest for the low-rank objects.

In the sector-wise architecture, we observe that generating symmetry
relations only from the local masters suffices to remove the redundancy. This
observation greatly simplifies the symmetry relations added to the global
reduction. Symmetry relations are generated in the settled round, after
the local masters of the sector are known. On F234 this produces $204$
relations. Generating from all seeds of the sector instead produces
$703{,}005$, while the final basis is the same $309$ masters in both cases.
The local masters are the lowest-complexity integrals of the sector, so the
$204$ relations are both fewer and individually cheaper than the $703{,}005$,
and the total saving is larger than the ratio of the counts suggests. The
sector-wise organization enables this saving, because the local masters of a
sector are known before the global reduction.

When a relation
involves an integral of another sector, that integral is registered as an
upstream target of that sector and the dependency is added to the scheduling
graph.

\subsection{Memory-aware scheduling}\label{subsec:sched}

Sector tasks are submitted to a thread pool of size \texttt{sthreads}. Two
scheduling modes are available. In \texttt{layer} mode the sectors are grouped
by popcount and a barrier separates consecutive layers, which guarantees the
dependency order and allows the upstream targets registered by a layer to be
merged before they are propagated, which removes duplicates. In \texttt{fine}
mode, the default, a sector waits only for its direct parents through an
atomic counter. The root sectors start immediately, and a finishing sector
decrements the counters of its children. The fine mode keeps more threads
occupied when sector run times differ widely.

Concurrency is limited by the measured memory of the process, not by a
per-sector estimate. Let $M_{\max}$ be the budget (\texttt{max\_memory}) and
$M_{\mathrm{res}}$ the reserve (\texttt{memory\_reserve}). The program reads
the resident set size $M_{\mathrm{used}}$ from
\texttt{/proc/self/\allowbreak status} and computes
\begin{equation}
  M_{\mathrm{avail}} = M_{\max} - M_{\mathrm{used}} .
  \label{eq:mavail}
\end{equation}
A task is admitted by a three-tier gate. It is admitted unconditionally if no
sector is running, and directly if $M_{\mathrm{avail}}>2M_{\mathrm{res}}$. In
the intermediate range it is admitted only if a binary flag $B$ is clear,
\begin{equation}
  M_{\mathrm{res}} < M_{\mathrm{avail}} \le 2M_{\mathrm{res}}
  \quad\wedge\quad B=0 .
  \label{eq:tightadmit}
\end{equation}
The flag records that a task has been admitted in this range, and a single
admission sets $B=1$, so that the free space released by one sector is taken
by one sector rather than by all waiting tasks at once. The flag is cleared when $M_{\mathrm{avail}}\le M_{\mathrm{res}}$
or when $M_{\mathrm{avail}}$ has dropped by more than $\Delta=10$\,GB from its
value at admission. The first tier excludes deadlock: if nothing runs, one
task is always admitted.

The reconstruction stage has its own budget, because each worker holds a
complete copy of the reduction store for the whole stage. After measuring the
store size $S_{\mathrm{store}}$ on a single-prime trial solve, the pool is
sized as
\begin{equation}
  k = \left\lfloor \frac{M_{\mathrm{ceil}}-M_{\mathrm{used}}}
  {S_{\mathrm{store}}} \right\rfloor ,
  \qquad
  M_{\mathrm{ceil}} = \min(M_{\max},M_{\mathrm{mach}})-M_{\mathrm{res}} ,
  \label{eq:reconpool}
\end{equation}
where $M_{\mathrm{mach}}$ is the total physical memory of the machine. The
pool is bounded by the number of cores and by one less than the number of
primes. On F1 at $n=1$, Eq.~\eqref{eq:reconpool} gives $48$ workers out of $64$, with a
measured $7.68$\,GB per worker inside a $420$\,GB budget. The run peaks at
$425.30$\,GB.

%% file: Sec4_Engine.tex
\section{Optimizations of the elimination engine}\label{sec:engine}

The sector stage decides which equations are retained and which integrals are
local masters. The global stage then solves the assembled linear system many
times, once for each pair of a prime and a sample point used in the
reconstruction.
Generically all these solves have the same rows and the same positions of
nonzero coefficients, and only the coefficient values change.
Sec.~\ref{subsec:plan} uses this repetition: the order of the elimination is
computed once and stored. Sec.~\ref{subsec:order} uses the freedom that
remains inside that order: rows with the same first unknown can be processed
in any order, and the choice changes how many terms the elimination moves.

\subsection{Elimination plan and replay}\label{subsec:plan}

Let the global system be $Ax=0$ with $m$ rows. The unknowns are the integrals
that occur in the retained equations, numbered as in
Sec.~\ref{subsec:notation}. A row is a list of terms, each term a pair of an
unknown number and a coefficient, and the row length is the number of terms.
The coefficients are rational functions of $d$ and the kinematic invariants.
Because they are reconstructed from values in finite fields, the same system
is solved at every pair of a prime and a sample point.

An elimination processes the rows in some order and solves one unknown per
row. The terms of a row are ordered by decreasing unknown number, and the
first unknown in this order is called the \emph{head} of the row. When a row
is reached, its first unknown that is not yet solved becomes its
\emph{pivot}. The row is multiplied by the inverse of the pivot coefficient,
so that the pivot has coefficient one, and the result is stored as the
solution of the pivot. A later row that contains an already solved unknown is
reduced by adding a multiple of the stored solution, which cancels that
unknown. This addition is a \emph{merge}. The order of the rows depends only on
the positions of the nonzero coefficients, not on their values.

The \emph{elimination plan} is the stored record of the structural part. For
each row $r$ it stores
\begin{equation}
  \pi(r)=\bigl(\,h(r),\ n(r),\ e(r)\,\bigr) ,
  \label{eq:plan-record}
\end{equation}
the head $h(r)$, the row length $n(r)$, and the equation number $e(r)$, which
is the position of the row in the generated system and makes the order
deterministic when all other entries are equal. The plan also stores the order
in which the rows are processed. It does not copy the terms: all rows are kept
in one contiguous array, the \emph{flat store}. Each row occupies a segment of
the array, and its terms are stored as consecutive pairs (unknown number,
coefficient) in the order stated above. The plan refers to a row by its
location in the array. Keeping all rows in one array rather than in one object
per row makes the memory access sequential.

The order is obtained by sorting the rows with a key that groups them by head,
so that rows with the same head are adjacent. The full key is given in
Sec.~\ref{subsec:order}. Given the order, the pivot of every row is fixed by
the rows before it, and so is the sequence of merges. The solved unknowns in a
row are the pivots of earlier rows that occur in it, and their order follows
from the stored order and the stored list. Replay therefore repeats neither
the search for the pivots nor the sorting of the rows at a new sample point.
The plan never reads a coefficient, so replay reproduces the direct
elimination.

To replay the plan, the worker writes the coefficient values of the current
point into the store and follows the stored order. For each row it takes the pivot
fixed by the order, merges into the row the stored solutions of its solved
unknowns, normalizes the row, and stores the result. After
the forward pass, the solutions are substituted back in reverse order, which
is the usual back substitution. Two details reduce the arithmetic. The inverse
of the pivot coefficient is computed once and applied to every coefficient of
the row, so the division in the normalization becomes a multiplication,
\begin{equation}
  \tilde a = a\cdot \mathrm{inv}_p(a_{\pi})\bmod p ,
  \label{eq:shoup}
\end{equation}
where $a_{\pi}$ is the pivot coefficient and $\mathrm{inv}_p$ is its inverse
modulo $p$ (Shoup precomputation~\cite{Shoup:2021arith}). A merge of two
sorted rows is linear in their lengths, and its result is again sorted.

The plan remains valid while the number of rows is unchanged and no
coefficient that the plan uses vanishes at the current point. The check is
conservative: a zero anywhere in the store, not only at a pivot, invalidates
the order and triggers a fresh computation from the nonzero entries. For
random coefficients the expected number of zeros is $N_t/p$, where $N_t$ is
the number of stored terms. At $p\simeq2^{31}$ this is $0.18$ for the F1
system, so the plan is reused for most primes and rebuilt for the rest. A
rebuild costs one sort of the row list and does not change the result.

The same plan and the same store serve every solve of the global stage: the
solves of the reconstruction at its primes and sample points, the smaller
re-solve that checks the selected equations against the targets, and the solve
that the symbolic backend performs before it reconstructs.

The plan was measured on F234 with symbolic $d$ in six runs that combine
three plan settings with the two states of the seed truncation. The settings
are the full plan, the plan without the recorded merges, which keeps the row
order and finds the pivots at each point, and no plan at all, which orders the
rows and finds the pivots at each point. The six runs return the same $309$
masters and byte-identical master and reduction files, so the plan changes the
cost and not the result. With the truncation off, the global stage takes
$1$m$14$s with the full plan, $1$m$44$s with the row order alone, and
$2$m$1$s with no plan. With the truncation on, the times are $1$m$55$s,
$2$m$26$s and $2$m$35$s. Most of the gain over the per-point elimination comes
from recording the merges, not only the order of the rows.

\subsection{Head score ordering}\label{subsec:order}

The number of equations selected for reconstruction directly affects memory,
but it is not simply positively correlated with the reconstruction time. In
the finite-field reconstruction of one family, fewer selected equations do not
always mean a faster reconstruction in practice, and a preliminary conclusion
is that the relevant quantity is the total number of terms moved during the
elimination at each sample point. We introduce a head score into the row
ordering to reduce that number.

The head fixes the first unknown of a row but not the order of rows that
share it. Several rows can share the same head, and they form a \emph{head
group}. A common choice orders the rows of a group by length, shortest first.
The head score below takes precedence over the length inside the group, which
remains as a tie-breaker. A row contains only unknowns whose number does not
exceed that of its head. Consider an unknown of the row that is the head
of another row. If it is smaller than the head of the current row, its row has
a smaller head and is processed earlier, so the unknown is solved by the time
the current row is reached and its solution is merged into the row. The head
of the current row is solved before the current row is processed only when
another row of the same group precedes it, and it is the pivot of the first
row of the group. The unknowns that enter a row through merges are the
\emph{fill-in} of the elimination. The
more head unknowns a row contains, the more merges it receives before it
reaches its pivot, and the more terms the elimination moves.

Let $H=\{h(r): r\ \text{nonempty}\}$ be the set of heads and let
$\mathcal{S}_r$ be the set of unknowns in row $r$. The \emph{head score} of
the row is
\begin{equation}
  \sigma(r)=\bigl|\mathcal{S}_r\cap H\bigr| .
  \label{eq:frontid}
\end{equation}
When the row is processed, every head unknown smaller than its own head has
already been solved and contributes one merge, and the head itself contributes
one unless the row is the first of its group. Unknowns outside $H$ may also be
solved before the row is reached, but they are not counted. A merge can
introduce further head unknowns through the stored solution, so $\sigma(r)$ is
an approximation rather than an exact count. It measures the part of the row
growth that can be predicted from the lists of unknowns before the elimination
starts, and a small score identifies a row that reaches its pivot after few
additions.

The rows are sorted by the tuple
\begin{equation}
  \Bigl(\ \mathbf{1}[r\ \text{empty}],\ h(r),\ \sigma(r),\ n(r),\
  \mathcal{L}(r),\ e(r)\ \Bigr) ,
  \label{eq:orderkey}
\end{equation}
compared lexicographically, where empty rows are placed last, $h(r)$ brings
the rows of a head group together, $\sigma(r)$ orders them inside the group
in increasing order, and the remaining three entries break ties. Here
$\mathcal{L}(r)$ is the list of the unknowns of row $r$ in the stored order.
The increasing score places the row with the fewest head unknowns first in its
group. That row becomes the stored solution of the shared head, so a small
first row keeps the solution that the other rows merge small. The score is
computed from the lists of unknowns and the heads alone, so it does not depend
on the coefficient values: it is used when the plan is built and recomputed
by the same rule when the plan is invalidated. It changes only the
order in which the rows of a head group are processed, not which integrals are
reducible.

On F234 with the truncation on, dropping $\sigma(r)$ from the ordering key
raises the selected set from $261{,}145$ to $270{,}633$ equations, the terms
moved per sample point from $5.35\times10^{8}$ to $6.68\times10^{8}$, and the
reconstruction time from $1$m$28$s to $1$m$53$s. With the truncation off the
score is neutral: it selects $555{,}833$ equations against $551{,}148$ and
moves $2.24\times10^{8}$ terms against $2.20\times10^{8}$, and the two runs
differ by about $1$\,s. The ordering is greedy: each row is placed by its own
score, without a search over the possible orders, so it is not optimal. It is
chosen because it is computable from the positions of the nonzero coefficients
alone and therefore adds no cost at replay time. Because the elimination
efficiency directly determines the reconstruction efficiency, this direction
merits further study.

%% file: Sec5_Install.tex
\section{Installation and usage}\label{sec:install}

\subsection{Building the program}

SecRed is written in C++ and builds with CMake. It requires GMP and MPFR,
FLINT for arbitrary-precision rational arithmetic, GiNaC for symbolic
manipulation, yaml-cpp for input parsing, jemalloc as the allocator, and two
external libraries that are linked at build time: FiniteFlow for modular
linear algebra and rational reconstruction, and FireFly for reconstruction of
symbolic systems. On Debian or Ubuntu the packaged dependencies are installed
with
\begin{lstlisting}[numbers=none]
sudo apt install build-essential cmake ninja-build libgmp-dev \
  libmpfr-dev libflint-dev libyaml-cpp-dev libginac-dev \
  libcln-dev libjemalloc-dev
\end{lstlisting}
FiniteFlow and FireFly are built from source and installed with their own
\texttt{configure} scripts. The SecRed source is cloned from the repository
listed in the Program Summary with
\begin{lstlisting}[numbers=none]
git clone https://github.com/chiyutuci/secred.git
cd secred
\end{lstlisting}
and the program is built with
\begin{lstlisting}[numbers=none]
cmake -S . -B build -DCMAKE_BUILD_TYPE=Release
cmake --build build -j
\end{lstlisting}
which produces \texttt{build/secred}. A dependency installed under a non-standard
prefix is passed with \texttt{-DCMAKE\_\allowbreak PREFIX\_\allowbreak PATH}. The build links
FiniteFlow and FireFly into the binary, so neither has to be on
\texttt{PATH} at run time.

\subsection{Input file and options}\label{subsec:options}

A run is started as \texttt{secred input.yaml}. The input has three blocks.
The \texttt{family} block gives the topology: its name, the space--time
dimension, the loop and external momenta, the kinematic invariants with their
mass dimensions, the scalar-product rules, the propagators as
\texttt{[momentum, mass]}, and optional cut flags. The \texttt{targets} block
lists the integrals to reduce, one index vector per line. The optional
\texttt{preferred} block names integrals to be kept as masters when the
mathematics allows it. It expresses a preference rather than a guarantee. A
minimal input for F234, with four of its $860$ benchmark targets, is

\noindent\begin{inputlisting}
\begin{lstlisting}[numbers=none]
family:
  name: F234
  dimension: d
  internals: [l1, l2]
  externals: [p1, p2, p3]
  sps_rules:
    - [p1, p1, 0]
    - [p2, p2, 0]
    - [p3, p3, 0]
    - [p1, p2, 1/3]
    - [p1, p3, 1/7]
    - [p2, p3, 9/2-1/3-1/7]
  propagators:
    - [l1, 3/2]
    - [l1-p2, 3/2]
    - [l1+p1/2-p2/2+p3/2, 0]
    - [l2, 3/2]
    - [l2+p3, 3/2]
    - [l2-p1/2-p2/2+p3/2, 0]
    - [l1+l2-p1/2-p2/2+p3/2, 3/2]
    - [l1+l2+p1/2-p2/2+p3/2, 3/2]
    - [l1+p1, 0]
settings:
  sthreads: 64          # sector threads
  gthreads: 64          # reconstruction threads
  parallel_mode: fine   # fine | layer
  ordering_mode: balance  # balance | normal
  subsector_filter_n: 1 # filter offset
  truncate: true        # seed truncation (probe_mode: targets by default)
  symmetry: true        # symmetry relations
  max_memory: 512       # GB
  memory_reserve: 30    # GB
  cache_dir: cache_F234 # output directory
targets:
  - [1, 1, 1, 1, 1, 0, 1, 1, -4]
  - [1, -2, 1, 1, 1, 1, 1, 1, -2]
  - [1, 3, 1, 1, 0, -2, 0, 1, 0]
  - [1, 1, 2, 0, 0, 1, 1, 1, -3]
\end{lstlisting}
\end{inputlisting}
The third target carries a double dot, the fourth a dot, and the first two
carry rank $4$.
The \texttt{settings} block is optional. The options below are grouped by
what they control, and each entry gives its default.

\emph{Parallelism and memory.}

\textbf{\texttt{sthreads}} (default $64$). Number of worker threads of the
sector stage.

\textbf{\texttt{gthreads}} (default \texttt{sthreads}). Number of concurrent
reconstructions in the global stage, i.e.\ the number of FireFly sample points
or FiniteFlow rational reconstructions.

\textbf{\texttt{parallel\_mode}} (default \texttt{fine}). In \texttt{fine}
mode a sector is scheduled as soon as its direct parents finish. In
\texttt{layer} mode the sectors of one popcount run as a batch: their upstream
targets are merged and deduplicated before propagation. Fine mode keeps more
threads occupied when sector run times differ widely.

\textbf{\texttt{max\_memory}} (default $512$\,GB). Memory budget of the
process.

\textbf{\texttt{memory\_reserve}} (default $30$\,GB). Headroom kept below
\texttt{max\_memory} for allocations outside the scheduler.

\textbf{\texttt{max\_primes}} (default $500$). Cap on the number of primes in
numeric reconstruction. The CRT stops at convergence, so the cap limits a run
rather than determining its length.

\textbf{\texttt{cache\_dir}} (default \texttt{cache}). Output directory.

\emph{Sector algorithms.}

\textbf{\texttt{subsector\_filter\_n}} (default $1$). Offset $n$ of
Eq.~\eqref{eq:silf}.

\textbf{\texttt{truncate}} (default \texttt{false}). The seed truncation of
Sec.~\ref{subsec:trunc}. When it is off, the whole seed box is admitted, no
truncation rule is derived and \texttt{seed\_level\_offset} is void. When it
is on, the caps are derived from the targets and the truncation probe defaults
to \texttt{targets}. Naming a \texttt{probe\_mode} also turns the truncation
on, with that scope. An explicit \texttt{truncate: false} overrides this
setting: it forces the unscoped \texttt{all} walk.

\textbf{\texttt{probe\_mode}} (default \texttt{targets} while the truncation
is on). Selects which refused masters may widen the box: \texttt{targets}
restricts the check to input targets, \texttt{all} lets every out-of-box
master widen the box. It is forced to \texttt{all} under
\texttt{truncate: false}.

\textbf{\texttt{seed\_level\_offset}} (default automatic). Overrides the
offset $o$ of Eq.~\eqref{eq:trunc-keep}. By default it is $0$ when all targets
are pure denominators and $1$ otherwise. It is ignored when the truncation is
off or \texttt{probe\_mode: all} is in effect.

\textbf{\texttt{symmetry}} (default \texttt{true}). Enables the master-based
symmetry relations.

\emph{Ordering.}

\textbf{\texttt{ordering\_mode}} (default \texttt{balance}). In
\texttt{balance} mode the seeds, the subsector columns and the global integral
numbering all use the balance key: total power first, then depth, rank, the
largest denominator power and the largest numerator power. The same comparator
is used at the three places, so the boundary, the filter and the elimination
agree on which integral is simpler. The \texttt{normal} mode replaces this
key by depth, then rank, with the index vector as the final tie-break, and
is intended for comparison.

\subsection{Output}\label{subsec:output}

All output is written below \texttt{cache\_dir}. \texttt{reductions} contains
the result: for each target, one line per term in the form
\texttt{[}$a_1,\dots,a_N$\texttt{]*coefficient}, with the target's own row
first (coefficient $-1$) followed by the masters, and a blank line between
targets. \texttt{masters\_sector} lists the master candidates left by the
sector stage, but this is not yet the final basis. \texttt{masters\_global} lists
the final basis and is written as soon as the global learn pass has fixed it,
before the reconstruction, which can be the longest phase of a run. The
directory \texttt{sectors/} holds the per-sector logs and target lists,
\texttt{secred.log} is a copy of the standard output, and
\texttt{firefly.log} is written on runs that used FireFly. Lines from the
FiniteFlow backend carry the \texttt{[fflow]} tag.

%% file: Sec6_Bench.tex
\section{Benchmarks}\label{sec:bench}

\subsection{Setup}\label{subsec:setup}

All runs were performed on a node with two AMD EPYC 7542 processors ($128$
logical threads), $512$\,GB of memory and Ubuntu 24.04, with the program built
by GCC 13.3 in release mode. The sector and global stages used $64$ threads
each in \texttt{fine} mode, $n=1$ and \texttt{balance} ordering unless stated
otherwise. The runs use the default \texttt{truncate: false} unless stated
otherwise, and the run with the truncation on is discussed in
Sec.~\ref{sec:f234}. The reported
equation counts are the \emph{raw} equations generated
by all settled sector systems, the equations forwarded \emph{to global}, and
the equations \emph{to reconstruction} selected from the global system. The
wall clock is the full run, and the peak memory is the resident set size
reported by the program.

The two families are shown in Fig.~\ref{fig:families}. F234 is the two-loop
five-point non-planar family of $J/\psi\to\gamma\gamma\gamma$ with $L=2$,
$N=9$, three independent external momenta and a massive leg $p_4^2=4m^2$. It
is a medium-size family, and current NNLO calculations routinely require
hundreds or thousands of such families per process. The $860$ targets have
$\max\Depth=8$ and $\max\Rank=4$, giving $D_{\max}=9$ and $R_{\max}=5$ over
$219$ non-trivial sectors. F1 is the four-loop four-point cut family of the
same process with $L=4$, $N=14$, $E=1$, $p^2=4m^2$ and the first three
propagators cut. Its $15{,}602$ targets have $\max\Depth=13$ and
$\max\Rank=3$, giving $D_{\max}=14$ and $R_{\max}=4$ over $860$ non-trivial
sectors. These bounds give $5.11\times10^{8}$ seeds without the corner cut.
F234 is small enough for the Laporta programs in the comparison to finish, and
F1 is large enough that a global Laporta run exhausts the node during equation
generation.

Table~\ref{tab:phases} splits the wall clock into the phases reported by the
program. The sector stage is a minority of the total in every run, and the
global stage dominates. This is why
Sec.~\ref{sec:engine} is devoted to the global elimination: the sector
algorithms decide how large the global system is, but most of the time is
spent in the global stage.

\begin{table}[htbp]
\centering
\caption{Phase timings of the benchmark runs. The
global stage includes the learn pass and the reconstruction. The F234
(targets) column uses
\texttt{truncate: true} with \texttt{probe\_mode: targets}, the F234 ($n=1$)
column the default \texttt{truncate: false}.}
\label{tab:phases}
\small
\setlength{\tabcolsep}{3pt}
\begin{tabular}{@{}lrrrr@{}}
\toprule
phase & F234 ($n=1$) & F234 (targets) & F1 ($n=0$) & F1 ($n=1$)\\
\midrule
preprocessing  & $<1$\,s & $<1$\,s & 1m17s & 6m52s\\
sector stage   & 10.1\,s & 2.2\,s & 10m16s & 1h24m58s\\
global stage   & 1m18s & 1m55s & 58m20s & 1h52m35s\\
\midrule
total          & 1m30s & 1m59s & 1h14m43s & 3h24m30s\\
peak memory    & 6.97\,GB & 8.42\,GB & 386.92\,GB & 425.30\,GB\\
\bottomrule
\end{tabular}
\par\smallskip
\raggedright\footnotesize
The phase timers are independent and do not add exactly to the total.
\end{table}

For F234 the propagators are, with $p_4=p_1+p_2+p_3$,
{\small
\begin{align}
  D_1 &= l_1^2-m^2, & D_2 &= (l_1-p_2)^2-m^2,\notag\\
  D_3 &= \Bigl(l_1+\tfrac{p_1-p_2+p_3}{2}\Bigr)^2, &
  D_4 &= l_2^2-m^2,\notag\\
  D_5 &= (l_2+p_3)^2-m^2, &
  D_6 &= \Bigl(l_2+\tfrac{-p_1-p_2+p_3}{2}\Bigr)^2,\notag\\
  D_7 &= \Bigl(l_1+l_2+\tfrac{-p_1-p_2+p_3}{2}\Bigr)^2-m^2, &
  D_8 &= \Bigl(l_1+l_2+\tfrac{p_1-p_2+p_3}{2}\Bigr)^2-m^2,\notag\\
  D_9 &= (l_1+p_1)^2 , & &
  \label{eq:f234}
\end{align}
}
with $p_1^2=p_2^2=p_3^2=0$, $p_4^2=4m^2$, $p_1\cdot p_2=s$,
$p_1\cdot p_3=t$ and $p_2\cdot p_3=2m^2-s-t$. The masses on
$D_1,D_2,D_4,D_5,D_7,D_8$ are equal. The numeric cross-check uses
$d=1999/500$, $s=1/3$, $t=1/7$ and $m=3/2$.

For F1 the propagators are, with $p^2=4m^2$,
{\small
\begin{align}
  D_1 &= l_3^2, & D_2 &= l_4^2,\notag\\
  D_3 &= (l_3+l_4-p)^2, & D_4 &= (l_2-l_3+p)^2-m^2,\notag\\
  D_5 &= l_1^2-m^2, & D_6 &= \Bigl(l_1+\tfrac{p}{2}\Bigr)^2,\notag\\
  D_7 &= l_2^2-m^2, & D_8 &= (l_1+l_2+p)^2,\notag\\
  D_9 &= \Bigl(l_1+l_2+\tfrac{p}{2}\Bigr)^2-m^2, &
  D_{10} &= \Bigl(l_3-\tfrac{p}{2}\Bigr)^2-m^2,\notag\\
  D_{11} &= (l_1+l_3)^2-m^2, &
  D_{12} &= \Bigl(l_4-\tfrac{p}{2}\Bigr)^2-m^2,\notag\\
  D_{13} &= (l_2+l_4)^2-m^2, & D_{14} &= (l_1+l_4)^2 ,
  \label{eq:f1}
\end{align}
}
with masses on $D_4,D_5,D_7,D_9,D_{10},D_{11},D_{12},D_{13}$, the first
three propagators cut, $m=1$ and $d=1999/500$. The input files for both
families are in the repository.

\begin{table*}[t]
\centering
\caption{F234 with symbolic $d$ and $860$ targets. ``Raw'' is the number of
generated equations, ``global'' the independent equations forwarded to the
global stage, and ``recon.'' the equations selected for reconstruction. The
two Kira~3.1 rows are private runs on the same node with $64$ threads, the
first with default settings and the second with per-sector truncation. The two
SecRed rows use $n=1$. The first has \texttt{truncate: false} and the second
\texttt{truncate: true} with \texttt{probe\_mode: targets}. The scans below use
\texttt{truncate: false}. Kira's peak is the high-water mark of its main
process, and FIRE's is the sampled resident peak of its process group, which
includes its worker processes.}
\label{tab:f234}
\small
\setlength{\tabcolsep}{3.2pt}
\begin{tabular}{@{}lrrrrrr@{}}
\toprule
program & raw & global & recon. & masters & wall & peak\\
\midrule
Kira 3.1        & 19{,}501{,}427 & 5{,}192{,}771 & 565{,}749 & 309 & 501.3\,s & 13.8\,GB\\
Kira 3.1 (trunc.) & 1{,}122{,}048 & 464{,}034 & 172{,}561 & 346 & 242.4\,s & 6.22\,GB\\
SecRed (trunc.\ off) & 14{,}478{,}997 & 1{,}162{,}358 & 555{,}833 & 309 & 90.1\,s & 6.97\,GB\\
SecRed (trunc.\ on)  & 1{,}113{,}377 & 414{,}834 & 261{,}145 & 309 & 119.0\,s & 8.42\,GB\\
FIRE 7.1 (sym.) & 3{,}177{,}403$^\ast$ & 2{,}525{,}345$^\ast$ & --- & 318 & 487\,s & 1.0\,GB\\
Blade           & --- & --- & --- & --- & 3600\,s$^\dagger$ & 33\,GB\\
NeatIBP         & --- & --- & --- & --- & 3600\,s$^\dagger$ & 25.6\,GB\\
\bottomrule
\end{tabular}
\par\smallskip
\raggedright\footnotesize
$^\ast$FIRE reports equations total/used, not the same three stages.
$^\dagger$Killed at the one-hour cap without producing coefficients, see the
text.
\end{table*}

\begin{widefigure}[t]
\centering
\begin{minipage}[b]{\benchwidth}
\centering
\begin{minipage}[b]{0.48\linewidth}\centering
  \includegraphics[width=\linewidth]{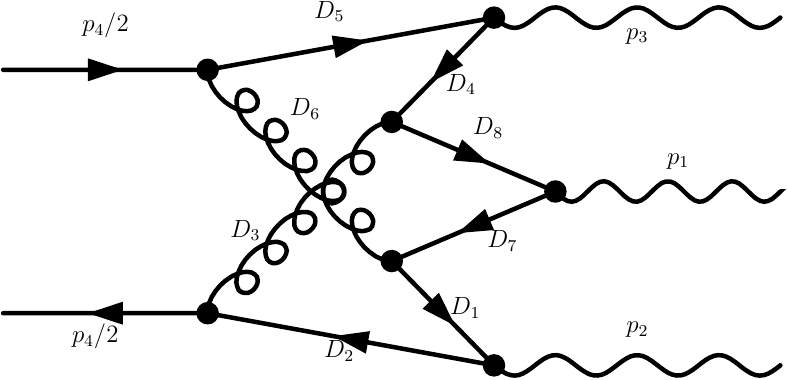}\\[-2pt]
  {\footnotesize (a) F234}
\end{minipage}\hfill
\begin{minipage}[b]{0.48\linewidth}\centering
  \includegraphics[width=\linewidth]{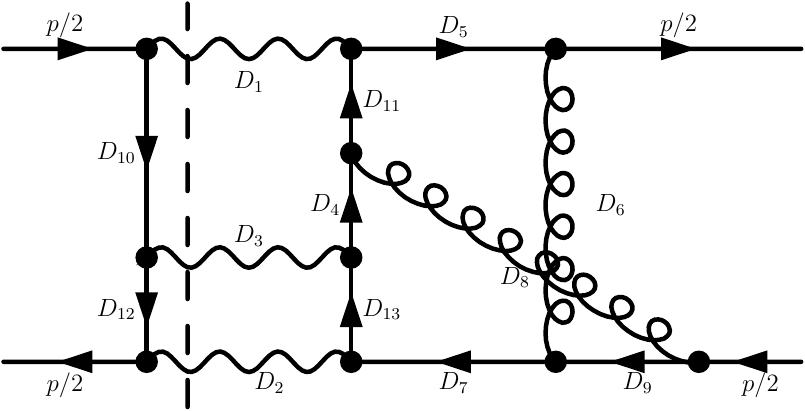}\\[-2pt]
  {\footnotesize (b) F1}
\end{minipage}
\end{minipage}
\caption{The two benchmark families. (a) F234: two loops, five points, nine
propagators. (b) F1: four loops, four points with a cut, fourteen
propagators.}
\label{fig:families}
\end{widefigure}

\subsection{Two-loop five-point non-planar family}\label{sec:f234}

Table~\ref{tab:f234} compares the programs on F234 with symbolic $d$. The
first SecRed row uses the adaptive seed box with its corner cut and no
truncation, and the second also enables the seed truncation, with
\texttt{probe\_mode: targets}. SecRed and default Kira return $309$ masters
each, but the bases differ: $250$ integrals are common and each program has
$59$ of its own, because the two programs use a different integral ordering.
The truncation-off run forwards $78\%$ fewer equations to the global stage,
uses half the peak memory and finishes in $90$\,s against $501$\,s. The number
of equations that reach reconstruction is essentially the same, so the gain is
not in the last stage: it is in the size of the global system and in the
memory needed to hold it.

Kira's per-sector truncation mode, the counterpart of the seed truncation of
Sec.~\ref{subsec:trunc}, cuts its own raw system to $1{,}122{,}048$ equations
and its global system to $464{,}034$, both below the $1{,}162{,}358$ forwarded
by the truncation-off SecRed run. The smaller system comes at a price. Kira's
basis grows to $346$ masters, a superset with $37$ extra integrals in proper
subsectors. The truncation shortens Kira's run from $501$\,s to $242$\,s and
lowers its peak from $13.8$\,GB to $6.22$\,GB, while the truncation-off SecRed
run takes $90$\,s. The per-sector truncation parameters follow the
same strategy as the Kira~3 interface of AMFlow~2.0~\cite{Huang:2026amflow2}.

The \texttt{targets} row of Table~\ref{tab:f234} restricts the correction to
masters on the reduction paths of the input targets. It returns the same $309$
masters and a byte-identical \texttt{reductions} file as the truncation-off
run, with $1{,}113{,}377$ raw equations instead of $14{,}478{,}997$. It takes $119$\,s
against $90$\,s for the truncation-off run. The reconstruction
receives fewer equations and yet takes $1$m$28$s, against $50$\,s in the
truncation-off run, so the smaller system does not translate into a shorter
reconstruction here. This is consistent with Sec.~\ref{subsec:order}, where
the reconstruction time followed the number of terms moved during the
elimination more closely than the number of equations. The truncation-on run
moves $5.35\times10^{8}$ terms per sample point, against $2.24\times10^{8}$ with the
truncation off. When the equation counts differ by less than one or two orders
of magnitude, fewer equations can correspond to more complicated reduction
paths, so the merges fill in more and move more terms. At $n=0$ the targets
mode returns $545$ masters: the narrower box accepts $236$ extra local
masters, the price anticipated in Sec.~\ref{subsec:trunc}, and they survive
into the final basis. The level filter at $n=1$ supplies the missing subsector
equations and restores the $309$-master basis. The truncation-off runs are the
conservative benchmark used for the scans below. The two programs therefore
use truncation to different ends: Kira accepts a larger basis for a shorter
run, while SecRed keeps the $309$-master basis and a smaller system but does
not gain time on this family.

To separate the cost of the equation system from the cost of the coefficients,
the same truncation-on run was repeated with $s$ and $t$ symbolic and $d$
fixed at $1999/500$. The equation system, the $309$ masters and the
\texttt{masters\_global} file are unchanged. The reconstruction is not: the
two-parameter run takes $2$h$23$m$13$s against minutes for the symbolic-$d$
counterpart, so the number of symbolic parameters sets the interpolation cost.
Running the same two-parameter input with \texttt{truncate: false} gives a
byte-identical \texttt{reductions} file and a global system larger by a
factor $2.8$, yet the run finishes in $1$h$38$m$25$s. The same inversion
persists: the truncation-on run moves $4.22\times10^{8}$ terms per sample point in
its merges and spends $1.843$\,s per sample point in forward elimination, the
truncation-off run $1.89\times10^{8}$ terms and $1.251$\,s.

FIRE~7.1 with its internal Lee symmetries and finite-field reconstruction
finishes the same target set in $487$\,s with $318$ masters, nine more than
SecRed and default Kira, and its process group peaks at $1.0$\,GB resident. The nine extra masters come from
the absence of a global stage: FIRE reduces each sector on its own, so an
integral whose elimination needs equations from several sectors stays in the
basis.

Blade and NeatIBP were each given a one-hour cap on the same input and did not
produce coefficients. Blade completed $10$ of its $11$ maximal-cut systems
before the cap. NeatIBP reported $100$ sectors finished and wrote master and
IBP files for $153$ sectors, but had not started the reduction when it was
stopped. These runs measure the cost of their preprocessing at this scale and
do not indicate a failure of the methods. Both are designed for families in
which their algebraic or numerical preprocessing is advantageous.

\begin{figure}[htbp]
\centering
\begin{tikzpicture}
\begin{axis}[
  width=0.48\columnwidth, height=0.36\columnwidth,
  symbolic x coords={0,1,2,3,$\infty$},
  xtick=data, xlabel={$n$}, ylabel={wall clock (s)},
  ymin=60, ymode=log, log basis y=10,
  tick label style={font=\scriptsize}, label style={font=\scriptsize},
  title style={font=\scriptsize}, title={(a)},
]
\addplot[mark=*,thick] coordinates {(0,70.53) (1,90.13) (2,100.13) (3,171.27) ({$\infty$},747.63)};
\end{axis}
\end{tikzpicture}\hfill
\begin{tikzpicture}
\begin{axis}[
  width=0.48\columnwidth, height=0.36\columnwidth,
  symbolic x coords={0,1,2,3,$\infty$},
  xtick=data, xlabel={$n$}, ylabel={peak memory (GB)},
  ymin=6, ymax=12,
  tick label style={font=\scriptsize}, label style={font=\scriptsize},
  title style={font=\scriptsize}, title={(b)},
]
\addplot[mark=square*,thick] coordinates {(0,6.68) (1,6.97) (2,7.56) (3,8.29) ({$\infty$},11.60)};
\end{axis}
\end{tikzpicture}
\caption{F234 as a function of the filter offset $n$: (a) wall clock on a
logarithmic scale and (b) peak memory. The basis is $314$ at $n=0$ and $309$
for all $n\ge1$.}
\label{fig:nscan}
\end{figure}
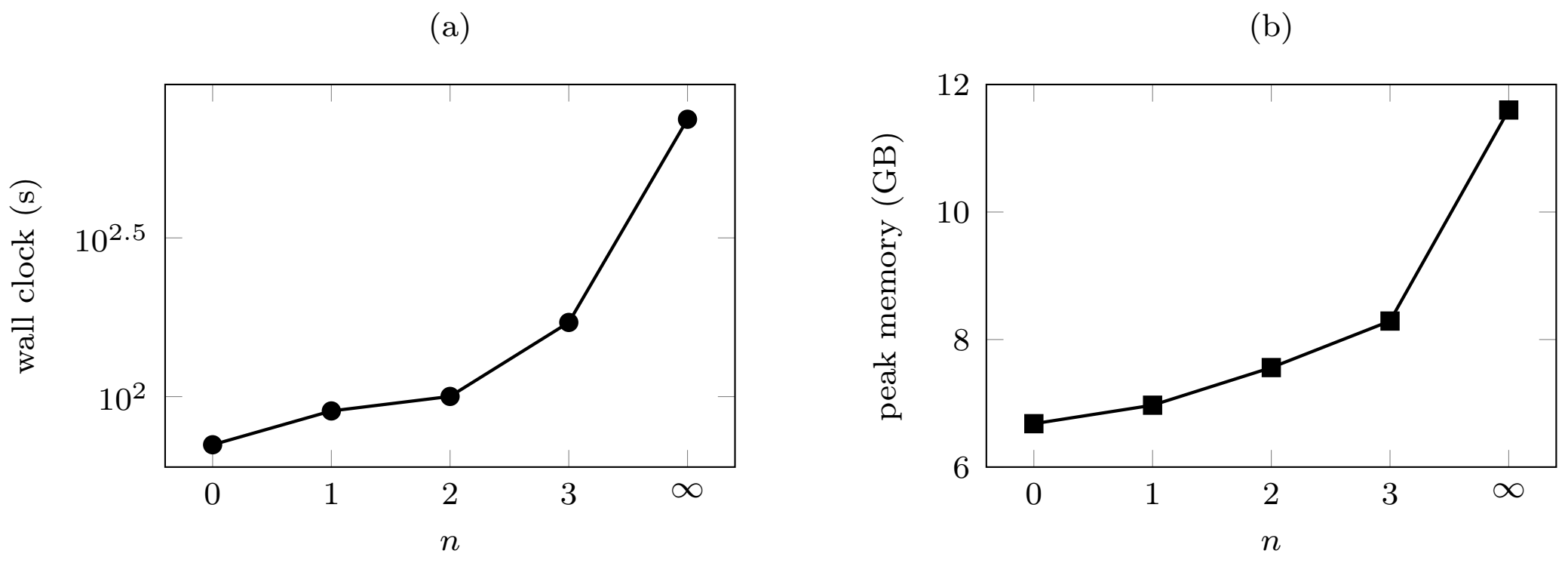

Table~\ref{tab:nscan} and Fig.~\ref{fig:nscan} scan the filter offset $n$. The
master count is $314$ at $n=0$ and $309$ for every $n\ge1$. The five extra
integrals at $n=0$ are reducible through cross-subsector relations that the
discarded columns would have supplied. Raising $n$ beyond $1$ does not change
the basis, and the raw equations stay nearly flat, but the global system grows
by a factor $3$ from $n=1$ to the unfiltered limit and the wall clock from
$1$m$30$s to $12$m$28$s. The extra subsector columns add few independent
sector equations. They add equations that only the global stage can use, and
the global elimination bears the cost. The default is $n=1$. The same scan on
F1 (Sec.~\ref{sec:f1}) has the opposite outcome, and is discussed there.

\begin{table}[htbp]
\centering
\caption{F234 as a function of the subsector filter offset $n$
(\texttt{balance}, symbolic $d$, $860$ targets, \texttt{truncate: false}).
The unfiltered limit is run with a finite $n$.}
\label{tab:nscan}
\small
\setlength{\tabcolsep}{3.6pt}
\begin{tabular}{@{}rrrrrrr@{}}
\toprule
$n$ & masters & raw eq. & to global & to recon. & wall & peak\\
\midrule
0     & 314 & 14{,}221{,}991 & 722{,}748 & 531{,}490 & 1m11s  & 6.68\,GB\\
1     & 309 & 14{,}478{,}997 & 1{,}162{,}358 & 555{,}833 & 1m30s   & 6.97\,GB\\
2     & 309 & 14{,}479{,}555 & 1{,}953{,}326 & 531{,}560 & 1m40s & 7.56\,GB\\
3     & 309 & 14{,}480{,}135 & 2{,}940{,}516 & 528{,}117 & 2m51s & 8.29\,GB\\
$\infty$ & 309 & 14{,}481{,}498 & 3{,}464{,}762 & 519{,}090 & 12m28s & 11.60\,GB\\
\bottomrule
\end{tabular}
\end{table}

The seed ordering is a separate choice from the filter offset.
Table~\ref{tab:defaults}(a) compares the balance key of
Sec.~\ref{subsec:notation} with the \texttt{normal} order on F234 at $n=1$. Both
return $309$ masters, but the balance order generates $44\%$ fewer global
equations and finishes in $1$m$30$s against $3$m$43$s, with $7.0$\,GB against
$12.1$\,GB. The balance order also lets the
truncation probe succeed in $208$ of the $219$ sectors at its first attempt, against
$157$ under the \texttt{normal} order, so fewer sectors incur the cost of a
discarded round.
The rest comes from the elimination
itself, which sees integrals in the same order in which the box and the filter
ranked them. The current version deliberately does not offer a user-defined
ordering of integral complexity, because the balance order is tightly coupled
to the sector-level algorithms: the level cap, the subsector filter and the
elimination all depend on it.

\begin{table*}[t]
\centering
\begin{minipage}{\defaultstablewidth}
\caption{Configuration studies behind the default settings. (a) F234 at $n=1$
with symbolic $d$ under the two seed orderings. (b) F1 with numerical
$d=1999/500$ and $15{,}602$ targets. The first $n=0$ row and the $n=1$ run
set \texttt{max\_memory} to $480$\,GB with a $60$\,GB reserve, a pool budget
of $420$\,GB. The second $n=0$ row sets \texttt{max\_memory} to $72$\,GB with
an $8$\,GB reserve, a pool budget of $64$\,GB. The Kira row uses per-sector
truncation set in its input.
``Raw'', ``global'' and ``recon.'' are as in
Table~\ref{tab:f234}.}
\label{tab:defaults}
\footnotesize
\setlength{\tabcolsep}{3.2pt}
\begin{tabular*}{\linewidth}{@{\extracolsep{\fill}}lrrrrrr@{}}
\toprule
 & masters & raw eq. & to global & to recon. & wall & peak\\
\midrule
\multicolumn{7}{@{}l}{(a) F234, $n=1$: seed ordering}\\
balance & 309 & 14{,}478{,}997 & 1{,}162{,}358 & 555{,}833 & 1m30s & 6.97\,GB\\
normal  & 309 & 19{,}053{,}931 & 2{,}083{,}375 & 927{,}520 & 3m43s & 12.10\,GB\\
\midrule
\multicolumn{7}{@{}l}{(b) F1: filter offset, Kira and FIRE}\\
$n=0$ & 566 & 1{,}105{,}719{,}427 & 34{,}557{,}830 & 29{,}590{,}539 & 1h14m43s & 386.92\,GB\\
$n=0$ (72\,GB) & 566 & 1{,}105{,}719{,}427 & 34{,}557{,}830 & 29{,}590{,}539 & 2h25m & 65\,GB\\
$n=1$ & 559 & 1{,}591{,}571{,}476 & 156{,}627{,}968 & 39{,}571{,}032 & 3h24m30s & 425.30\,GB\\
Kira (trunc.) & 703 & 108{,}139{,}890 & 39{,}118{,}736 & 5{,}865{,}036 & 1h40m & 37\,GB\\
FIRE 7.1 (cut) & 568 & 159{,}361{,}334$^\ast$ & 126{,}295{,}720$^\ast$ & --- & 4h06m & 26\,GB\\
\bottomrule
\end{tabular*}
\par\smallskip
\raggedright\footnotesize
$^\ast$FIRE reports generated and used equations, not the same stages as the
other rows.
\end{minipage}
\end{table*}

\subsection{Four-loop four-point cut family}\label{sec:f1}

Table~\ref{tab:defaults}(b) reports F1 at $n=0$ and $n=1$, with a Kira run
using per-sector truncation and an analytic FIRE~7.1 run. A Kira run without
truncation exhausts the node's memory during equation generation, so the Kira
comparison uses truncation. SecRed returns $566$ masters at $n=0$ and $559$ at
$n=1$. FIRE returns $568$ and Kira $703$, so SecRed gives the smallest basis
on this family.

At $n=0$ the run finishes in $1$h$14$m$43$s, while FIRE runs for $4$h$06$m
and Kira for $1$h$40$m. With \texttt{max\_memory} set to $72$\,GB and
\texttt{memory\_reserve} to $8$\,GB, the pool budget is $64$\,GB, and the
same run takes $2$h$25$m, between the Kira and FIRE times. At $n=1$ the basis has
$7$ fewer masters than at $n=0$ and is a strict subset of it. The cost is
substantial: the raw equations grow by $44\%$, the global system by a factor
$4.5$, and the wall clock by a factor $2.7$. For this family the extra
completeness does not justify this cost, and the production setting is $n=0$.

Because most of the memory is used in the reconstruction stage, the peak is
controlled by the size of the reconstruction pool, which the scheduler derives
from \texttt{max\_memory} and the reserve. At $n=0$ \texttt{max\_memory} is
$480$\,GB; with the $60$\,GB reserve, the pool budget is $420$\,GB, the pool
runs $64$ of $64$ workers at $5.72$\,GB per worker, and the peak is
$386.92$\,GB. With \texttt{max\_memory} at $72$\,GB and the reserve at
$8$\,GB, the budget is $64$\,GB, the pool holds eight workers, and the peak
drops to $65$\,GB. At $n=1$ Eq.~\eqref{eq:reconpool} sizes the pool to $48$
of $64$ workers, with a measured $7.68$\,GB per worker inside the same
budget. The run peaks at $425.30$\,GB, $5.3$\,GB above
the budget, and the $60$\,GB reserve absorbs the overshoot. Without the pool,
the first attempt at this configuration ran out of memory while filling the
reconstruction workers. Sizing the pool from the measured store allows the
run to fit within the budget. The thread count therefore needs no manual
tuning: the user sets only the maximum thread count, \texttt{max\_memory} and
\texttt{memory\_reserve}, and the scheduler sizes the pool to the largest
number of workers that fits the budget.

%% file: Sec7_Summary.tex
\section{Summary}\label{sec:summary}

We present SecRed 1.0, a program that reduces Feynman integrals
sector by sector. Each sector is processed independently and only its
linearly independent equations are forwarded to a global stage, so that the
working set of a sector remains local and the sectors form independent tasks
for dynamic scheduling under a memory budget.

The sector stage combines an adaptive seed box with a corner cut, a seed
truncation with self-correcting caps, a subsector level filter and symmetry
relations expanded only from the local masters. The box and the truncation
widen or relax themselves when the settled basis shows that they removed too
much. The result is a substantially smaller system entering the elimination,
with the self-correction preserving the completeness of the basis and the
symmetry relations removing most of the expansions. The global stage records
the elimination plan once and replays it at every prime and sample point, so
that only the arithmetic is repeated, while the head score orders the rows of
a head group so that the merges move fewer terms. These design choices and
optimizations, combined with the memory control, allow the program to run
efficiently within a limited memory budget.

The correctness of the reductions has been cross-checked against independent
implementations. SecRed serves as the IBP engine of AMFlow~\cite{Liu:2022chg}
in the NNLO corrections to $J/\psi\to\gamma\gamma\gamma$~\cite{Zeng:2026jpsi3gamma},
in double $J/\psi$ production at $B$ factories~\cite{Huang:2024doublejpsi},
and in an ongoing $J/\psi\to ggg$ calculation. The program is expected to
support deeper and more demanding calculations in high-energy physics in the
years ahead.

Several directions remain for improvement. The seed set remains the main
source of avoidable cost. The adaptive box is conservative and the truncation
is aggressive. Although both adjust themselves when a needed integral is
missed, a scheme that is smaller while remaining safe, in the sense that it
retains the information required for the minimal master basis, has yet to be
found. Recent proposals point in this direction: a tube of seeds along
the path between a target and its masters~\cite{Berman:2026tube}, and
formulations that avoid the seed set entirely~\cite{delaCruz:2026seedless}.

The size of the system, however, is not the only factor in the run time. A
smaller system is beneficial, but it does not necessarily accelerate the
elimination. The run time depends more strongly on the reduction paths and on
the fill-in of the
merges than on the number of equations, and the mechanism behind this
relation is not yet sufficiently understood to guide the choice of the
equation set. Machine-learning methods explore the structure of the equation
set by searching for heuristics that decide which equations enter the
reduction~\cite{vonHippel:2025ml,Song:2026funsearch}. The elimination
algorithm itself admits further optimization, and the finite-field
reconstruction stage also offers scope for improvement, for example through
GPU acceleration.

SecRed is designed for large-scale, batch automated reduction, such as the
numerical evaluation of master integrals with AMFlow, where many families are
reduced in a single campaign. The aim is to achieve efficiency without manual
tuning and stable behavior across the batch, and the benchmarks are selected
accordingly.
For particular mathematical structures or special requirements, for
calculations that permit manual parameter tuning, or for reconstructions with
many symbolic parameters, further optimization remains possible.

%% file: Sec8_Contrib.tex
\section*{Author contributions}

Ruichang Niu designed the algorithms, implemented and optimized the program,
and wrote the manuscript. Chao Zeng tested the program and provided the
benchmark cases. Bin Gong and Jianxiong Wang contributed to the discussion of
the algorithm design and coordinated the work. All authors reviewed the
manuscript. Generative AI tools were used in code development and for polishing
drafts of the manuscript. The authors reviewed the output and take full
responsibility for the content of the publication.

%% file: sec.bib
@article{Tkachov:1981wb,
  author  = {F. V. Tkachov},
  title   = {A theorem on analytical calculability of four loop renormalization group functions},
  journal = {Phys. Lett. B},
  volume  = {100},
  pages   = {65--68},
  year    = {1981}
}

@article{Chetyrkin:1981qh,
  author  = {K. G. Chetyrkin and F. V. Tkachov},
  title   = {Integration by parts: the algorithm to calculate $\beta$-functions in 4 loops},
  journal = {Nucl. Phys. B},
  volume  = {192},
  pages   = {159--204},
  year    = {1981}
}

@article{Gehrmann:1999as,
  author  = {T. Gehrmann and E. Remiddi},
  title   = {Differential equations for two-loop four-point functions},
  journal = {Nucl. Phys. B},
  volume  = {580},
  pages   = {485--518},
  year    = {2000},
  eprint  = {hep-ph/9912329}
}

@article{Laporta:2000dsw,
  author  = {S. Laporta},
  title   = {High-precision calculation of multiloop {Feynman} integrals by difference equations},
  journal = {Int. J. Mod. Phys. A},
  volume  = {15},
  pages   = {5087--5159},
  year    = {2000},
  eprint  = {hep-ph/0102033}
}

@article{Tarasov:1996,
  author  = {O. V. Tarasov},
  title   = {Connection between {Feynman} integrals having different values of the space-time dimension},
  journal = {Phys. Rev. D},
  volume  = {54},
  pages   = {6479--6490},
  year    = {1996},
  eprint  = {hep-th/9606018}
}

@article{Lee:2010aa,
  author  = {R. N. Lee},
  title   = {Space-time dimensionality $D$ as complex variable: calculating loop integrals using dimensional recurrence relation and analytical properties with respect to $D$},
  journal = {Nucl. Phys. B},
  volume  = {830},
  pages   = {474--492},
  year    = {2010},
  eprint  = {0911.0252}
}

@article{Smirnov:2011num,
  author  = {A. V. Smirnov and A. V. Petukhov},
  title   = {The number of master integrals is finite},
  journal = {Lett. Math. Phys.},
  volume  = {97},
  pages   = {37--44},
  year    = {2011},
  eprint  = {1004.4199}
}

@article{Grozin:2011,
  author  = {A. G. Grozin},
  title   = {Integration by parts: an introduction},
  journal = {Int. J. Mod. Phys. A},
  volume  = {26},
  pages   = {2807},
  year    = {2011},
  eprint  = {1104.3993}
}

@article{Anastasiou:2004vj,
  author  = {C. Anastasiou and A. Lazopoulos},
  title   = {Automatic integral reduction for higher order perturbative calculations},
  journal = {JHEP},
  volume  = {07},
  pages   = {046},
  year    = {2004},
  eprint  = {hep-ph/0404258}
}

@article{Studerus:2009ye,
  author  = {C. Studerus},
  title   = {Reduze -- {Feynman} integral reduction in {C++}},
  journal = {Comput. Phys. Commun.},
  volume  = {181},
  pages   = {1293--1300},
  year    = {2010},
  eprint  = {0912.2546}
}

@article{vonManteuffel:2012np,
  author  = {A. von Manteuffel and C. Studerus},
  title   = {Reduze 2 -- {Distributed} {Feynman} {Integral} {Reduction}},
  journal = {\href{https://arxiv.org/abs/1201.4330}{arXiv:1201.4330 [hep-ph]}},
  year    = {2012}
}

@article{Smirnov:2008iw,
  author  = {A. V. Smirnov},
  title   = {Algorithm {FIRE} -- {Feynman} {Integral} {REduction}},
  journal = {JHEP},
  volume  = {10},
  pages   = {107},
  year    = {2008},
  eprint  = {0807.3243}
}

@article{Smirnov:2019qkx,
  author  = {A. V. Smirnov and F. S. Chukharev},
  title   = {{FIRE6}: Feynman {Integral} {REduction} with modular arithmetic},
  journal = {Comput. Phys. Commun.},
  volume  = {247},
  pages   = {106877},
  year    = {2020},
  eprint  = {1901.07808}
}

@article{Lee:2013mka,
  author  = {R. N. Lee},
  title   = {Presenting {LiteRed}: a tool for the {Loop} {InTEgrals} {REDuction}},
  journal = {\href{https://arxiv.org/abs/1212.2685}{arXiv:1212.2685 [hep-ph]}},
  year    = {2012}
}

@article{Lee:2014tja,
  author  = {R. N. Lee},
  title   = {{LiteRed} 1.4: a powerful tool for reduction of multiloop integrals},
  journal = {J. Phys. Conf. Ser.},
  volume  = {523},
  pages   = {012059},
  year    = {2014},
  eprint  = {1310.1145}
}

@article{Kant:2013vta,
  author  = {P. Kant},
  title   = {Finding linear dependencies in integration-by-parts equations: a {Monte} {Carlo} approach},
  journal = {Comput. Phys. Commun.},
  volume  = {185},
  pages   = {1473--1476},
  year    = {2014},
  eprint  = {1309.7287}
}

@article{vonManteuffel:2014ixa,
  author  = {A. von Manteuffel and R. M. Schabinger},
  title   = {A novel approach to integration by parts reduction},
  journal = {Phys. Lett. B},
  volume  = {744},
  pages   = {101--104},
  year    = {2015},
  eprint  = {1406.4513}
}

@article{Peraro:2016wsq,
  author  = {T. Peraro},
  title   = {Scattering amplitudes over finite fields and multivariate functional reconstruction},
  journal = {JHEP},
  volume  = {12},
  pages   = {030},
  year    = {2016},
  eprint  = {1608.01902}
}

@article{Peraro:2019sve,
  author  = {T. Peraro},
  title   = {{FiniteFlow}: multivariate functional reconstruction using finite fields and dataflow graphs},
  journal = {JHEP},
  volume  = {07},
  pages   = {031},
  year    = {2019},
  eprint  = {1905.08019}
}

@article{Klappert:2020firefly,
  author  = {J. Klappert and F. Lange},
  title   = {Reconstructing rational functions with {FireFly}},
  journal = {Comput. Phys. Commun.},
  volume  = {247},
  pages   = {106951},
  year    = {2020},
  eprint  = {1904.00009}
}

@article{Klappert:2021int,
  author  = {J. Klappert and S. Y. Klein and F. Lange},
  title   = {Interpolation of dense and sparse rational functions and other improvements in {FireFly}},
  journal = {Comput. Phys. Commun.},
  volume  = {264},
  pages   = {107968},
  year    = {2021},
  eprint  = {2004.01463}
}

@article{Belitsky:2023,
  author  = {A. V. Belitsky and A. V. Smirnov and R. V. Yakovlev},
  title   = {Balancing act: multivariate rational reconstruction for {IBP}},
  journal = {Nucl. Phys. B},
  volume  = {993},
  pages   = {116253},
  year    = {2023},
  eprint  = {2303.02511}
}

@article{Chawdhry:2024,
  author  = {H. A. Chawdhry},
  title   = {p-adic reconstruction of rational functions in multiloop amplitudes},
  journal = {Phys. Rev. D},
  volume  = {110},
  pages   = {056028},
  year    = {2024},
  eprint  = {2312.03672}
}

@inproceedings{Zippel:1979,
  author    = {R. Zippel},
  title     = {Probabilistic algorithms for sparse polynomials},
  booktitle = {Symbolic and Algebraic Computation (EUROSAM '79)},
  editor    = {E. W. Ng},
  pages     = {216--226},
  year      = {1979}
}

@article{Zippel:1990,
  author  = {R. Zippel},
  title   = {Interpolating polynomials from their values},
  journal = {J. Symb. Comput.},
  volume  = {9},
  pages   = {375--403},
  year    = {1990}
}

@inproceedings{Monagan:2004,
  author    = {M. Monagan},
  title     = {Maximal quotient rational reconstruction: an almost optimal algorithm for rational reconstruction},
  booktitle = {Proceedings of the 2004 International Symposium on Symbolic and Algebraic Computation (ISSAC '04)},
  pages     = {243--249},
  year      = {2004}
}

@article{Maierhoefer:2017hyi,
  author  = {P. Maierh\"{o}fer and J. Usovitsch and P. Uwer},
  title   = {Kira -- {A} {Feynman} integral reduction program},
  journal = {Comput. Phys. Commun.},
  volume  = {230},
  pages   = {99--112},
  year    = {2018},
  eprint  = {1705.05610}
}

@article{Klappert:2020nbg,
  author  = {J. Klappert and F. Lange and P. Maierh\"{o}fer and J. Usovitsch},
  title   = {Integral reduction with {Kira} 2.0 and finite field methods},
  journal = {Comput. Phys. Commun.},
  volume  = {266},
  pages   = {108024},
  year    = {2021},
  eprint  = {2008.06494}
}

@article{Lange:2025fba,
  author  = {F. Lange and J. Usovitsch and Z. Wu},
  title   = {Kira 3: Integral reduction with efficient seeding and optimized equation selection},
  journal = {Comput. Phys. Commun.},
  volume  = {322},
  pages   = {109999},
  year    = {2026}
}

@article{Smirnov:2025fire7,
  author  = {A. V. Smirnov and M. Zeng},
  title   = {{FIRE} 7: Automatic reduction with modular approach},
  journal = {Comput. Phys. Commun.},
  volume  = {325},
  pages   = {110200},
  year    = {2026},
  doi     = {10.1016/j.cpc.2026.110200},
  eprint  = {2510.07150}
}

@article{Liu:2022chg,
  author  = {X. Liu and Y.-Q. Ma},
  title   = {{AmFlow}: A {Mathematica} package for {Feynman} integrals computation via auxiliary mass flow},
  journal = {Comput. Phys. Commun.},
  volume  = {283},
  pages   = {108565},
  year    = {2023},
  eprint  = {2201.11669}
}

@article{Wu:2023upw,
  author  = {Z. Wu and J. B\"{o}hm and R. Ma and H. Xu and Y. Zhang},
  title   = {{NeatIBP} 1.0, a package generating small-size integration-by-parts relations for {Feynman} integrals},
  journal = {Comput. Phys. Commun.},
  volume  = {295},
  pages   = {108999},
  year    = {2024},
  eprint  = {2305.08783}
}

@article{Wu:2025aen,
  author  = {Z. Wu and J. B\"{o}hm and R. Ma and J. Usovitsch and Y. Xu and Y. Zhang},
  title   = {Performing integration-by-parts reductions using {NeatIBP} 1.1 + {Kira}},
  journal = {Comput. Phys. Commun.},
  volume  = {316},
  pages   = {109798},
  year    = {2025}
}

@article{Guan:2025blade,
  author  = {X. Guan and X. Liu and Y.-Q. Ma and W.-H. Wu},
  title   = {Blade: A package for block-triangular form improved {Feynman} integrals decomposition},
  journal = {Comput. Phys. Commun.},
  volume  = {310},
  pages   = {109538},
  year    = {2025}
}

@article{Georgoudis:2017azurite,
  author  = {A. Georgoudis and K. J. Larsen and Y. Zhang},
  title   = {Azurite: An algebraic geometry based package for finding bases of loop integrals},
  journal = {Comput. Phys. Commun.},
  volume  = {221},
  pages   = {203--215},
  year    = {2017},
  eprint  = {1612.04252}
}

@article{Gluza:2011,
  author  = {J. Gluza and K. Kajda and D. A. Kosower},
  title   = {Towards a basis for planar two-loop integrals},
  journal = {Phys. Rev. D},
  volume  = {83},
  pages   = {045012},
  year    = {2011},
  eprint  = {1009.0472}
}

@article{Schabinger:2012,
  author  = {R. M. Schabinger},
  title   = {A new algorithm for the generation of unitarity-compatible integration by parts relations},
  journal = {JHEP},
  volume  = {01},
  pages   = {077},
  year    = {2012},
  eprint  = {1111.4220}
}

@article{Bohm:2017xot,
  author  = {J. B\"{o}hm and A. Georgoudis and K. J. Larsen and M. Schulze and Y. Zhang},
  title   = {Complete sets of logarithmic vector fields for integration-by-parts identities of {Feynman} integrals},
  journal = {Phys. Rev. D},
  volume  = {98},
  pages   = {025023},
  year    = {2018},
  eprint  = {1712.09737}
}

@article{Bohm:2018hexagonbox,
  author  = {J. B\"{o}hm and A. Georgoudis and K. J. Larsen and H. Sch\"{o}nemann and Y. Zhang},
  title   = {Complete integration-by-parts reductions of the non-planar hexagon-box via module intersections},
  journal = {JHEP},
  volume  = {09},
  pages   = {024},
  year    = {2018},
  eprint  = {1805.01873}
}

@article{Kosower:2018,
  author  = {D. A. Kosower},
  title   = {Direct solution of integration-by-parts systems},
  journal = {Phys. Rev. D},
  volume  = {98},
  pages   = {025008},
  year    = {2018},
  eprint  = {1804.00131}
}

@article{Larsen:2015ped,
  author  = {K. J. Larsen and Y. Zhang},
  title   = {Integration-by-parts reductions from unitarity cuts and algebraic geometry},
  journal = {Phys. Rev. D},
  volume  = {93},
  pages   = {041701},
  year    = {2016},
  eprint  = {1511.01071}
}

@article{Baikov:2006,
  author  = {P. A. Baikov},
  title   = {A practical criterion of irreducibility of multi-loop {Feynman} integrals},
  journal = {Phys. Lett. B},
  volume  = {634},
  pages   = {325--329},
  year    = {2006},
  eprint  = {hep-ph/0507053}
}

@article{Lee:2008grp,
  author  = {R. N. Lee},
  title   = {Group structure of the integration-by-part identities and its application to the reduction of multiloop integrals},
  journal = {JHEP},
  volume  = {07},
  pages   = {031},
  year    = {2008},
  eprint  = {0804.3008}
}

@article{Pak:2011xt,
  author  = {A. Pak},
  title   = {The toolbox of modern multi-loop calculations: novel analytic and semi-analytic techniques},
  journal = {J. Phys. Conf. Ser.},
  volume  = {368},
  pages   = {012049},
  year    = {2012},
  eprint  = {1111.0868}
}

@article{Wu:2025symmetry,
  author  = {Z. Wu and Y. Zhang},
  title   = {A new method for finding more symmetry relations of {Feynman} integrals},
  journal = {Comput. Phys. Commun.},
  volume  = {314},
  pages   = {109681},
  year    = {2025},
  eprint  = {2406.20016}
}

@book{Nakanishi:1971,
  author    = {N. Nakanishi},
  title     = {Graph theory and Feynman integrals},
  publisher = {Gordon and Breach},
  year      = {1971}
}

@article{Bogner:2010,
  author  = {C. Bogner and S. Weinzierl},
  title   = {Feynman graph polynomials},
  journal = {Int. J. Mod. Phys. A},
  volume  = {25},
  pages   = {2585--2618},
  year    = {2010},
  eprint  = {1002.3458}
}

@article{Smirnov:2020cho,
  author  = {A. V. Smirnov and V. A. Smirnov},
  title   = {How to choose master integrals},
  journal = {Nucl. Phys. B},
  volume  = {960},
  pages   = {115213},
  year    = {2020},
  eprint  = {2002.08042}
}

@article{Boehm:2020simple,
  author  = {J. Boehm and M. Wittmann and Z. Wu and Y. Xu and Y. Zhang},
  title   = {{IBP} reduction coefficients made simple},
  journal = {JHEP},
  volume  = {12},
  pages   = {054},
  year    = {2020},
  eprint  = {2008.13194}
}

@article{Kotikov:1991,
  author  = {A. V. Kotikov},
  title   = {Differential equations method: new technique for massive {Feynman} diagrams calculation},
  journal = {Phys. Lett. B},
  volume  = {254},
  pages   = {158--164},
  year    = {1991}
}

@article{Henn:2013,
  author  = {J. M. Henn},
  title   = {Multiloop integrals in dimensional regularization made simple},
  journal = {Phys. Rev. Lett.},
  volume  = {110},
  pages   = {251601},
  year    = {2013},
  eprint  = {1304.1806}
}

@article{Lee:2015reducingde,
  author  = {R. N. Lee},
  title   = {Reducing differential equations for multiloop master integrals},
  journal = {JHEP},
  volume  = {04},
  pages   = {108},
  year    = {2015},
  eprint  = {1411.0911}
}

@article{Bauer:2000cp,
  author  = {C. W. Bauer and A. Frink and R. Kreckel},
  title   = {Introduction to the {GiNaC} framework for symbolic computation within the {C++} programming language},
  journal = {J. Symb. Comput.},
  volume  = {33},
  pages   = {1--12},
  year    = {2002},
  eprint  = {cs/0004015}
}

@misc{flint,
  author  = {W. B. Hart and F. Johansson and S. Pancratz},
  title   = {{FLINT}: Fast {Library} for {Number} {Theory}},
  howpublished = {\url{https://flintlib.org}}
}

@misc{yamlcpp,
  author  = {J. Beder and others},
  title   = {yaml-cpp: A {YAML} parser and emitter for {C++}},
  howpublished = {\url{https://github.com/jbeder/yaml-cpp}}
}

@article{Shoshany:2024threadpool,
  author  = {B. Shoshany},
  title   = {A {C++17} thread pool for high-performance scientific computing},
  journal = {SoftwareX},
  volume  = {26},
  pages   = {101687},
  year    = {2024},
  doi     = {10.1016/j.softx.2024.101687},
  eprint  = {2105.00613}
}

@misc{Vermaseren:2000nd,
  author  = {J. A. M. Vermaseren},
  title   = {New features of {FORM}},
  howpublished = {\url{https://arxiv.org/abs/math-ph/0010025}},
  year    = {2000}
}

@article{Zeng:2026jpsi3gamma,
  author  = {C. Zeng and B. Gong and J.-X. Wang and R.-C. Niu and X.-D. Huang and C. Li},
  title   = {Complete next-to-next-to-leading-order {QCD} correction to {$J/\psi\to 3\gamma$} decay},
  journal = {\href{https://arxiv.org/abs/2603.26199}{arXiv:2603.26199 [hep-ph]}},
  year    = {2026}
}

@article{Huang:2024doublejpsi,
  author  = {X.-D. Huang and B. Gong and R.-C. Niu and H.-M. Yu and J.-X. Wang},
  title   = {Next-to-next-to-leading-order {QCD} corrections to double {$J/\psi$} production at the {$B$} factories},
  journal = {JHEP},
  volume  = {02},
  pages   = {055},
  year    = {2024},
  eprint  = {2311.04751}
}

@article{Abreu:2019fivegluon,
  author  = {S. Abreu and J. Dormans and F. Febres Cordero and H. Ita and B. Page},
  title   = {Analytic Form of Planar Two-Loop Five-Gluon Scattering Amplitudes in {QCD}},
  journal = {Phys. Rev. Lett.},
  volume  = {122},
  pages   = {082002},
  year    = {2019},
  eprint  = {1812.04586}
}

@article{Anastasiou:2015higgs,
  author  = {C. Anastasiou and C. Duhr and F. Dulat and F. Herzog and B. Mistlberger},
  title   = {Higgs Boson Gluon-Fusion Production in {QCD} at Three Loops},
  journal = {Phys. Rev. Lett.},
  volume  = {114},
  pages   = {212001},
  year    = {2015},
  eprint  = {1503.06056}
}

@article{Duhr:2020dy,
  author  = {C. Duhr and F. Dulat and B. Mistlberger},
  title   = {Charged current {Drell-Yan} production at {N}$^3${LO}},
  journal = {JHEP},
  volume  = {11},
  pages   = {143},
  year    = {2020},
  eprint  = {2007.13313}
}

@article{Huss:2026wishlist,
  author  = {A. Huss and J. Huston and S. Jones and M. Pellen and R. R{\"o}ntsch},
  title   = {Les Houches 2023 -- Physics at {TeV} colliders: Report on the Standard Model precision wishlist},
  journal = {SciPost Phys. Comm. Rep.},
  volume  = {21},
  year    = {2026},
  eprint  = {2504.06689},
  archivePrefix = {arXiv},
  doi     = {10.21468/SciPostPhysCommRep.21}
}

@inproceedings{Papadopoulos:2025hoc,
  author    = {C. G. Papadopoulos},
  title     = {Higher-Order Corrections at the {LHC}: Current Status and Prospects},
  booktitle = {Springer Proceedings in Physics},
  year      = {2025},
  pages     = {49--59},
  publisher = {Springer}
}

@article{Smirnov:2025review,
  author  = {A. V. Smirnov and V. A. Smirnov},
  title   = {Two decades of algorithmic {Feynman} integral reduction},
  journal = {Numer. Methods Program.},
  volume  = {26},
  pages   = {429},
  year    = {2025},
  eprint  = {2510.10748},
  archivePrefix = {arXiv},
  doi     = {10.26089/NumMet.v26r429}
}

@incollection{Shoup:2021arith,
  author    = {V. Shoup},
  title     = {Arithmetic software libraries},
  booktitle = {Computational Cryptography},
  editor    = {J. Bos and M. Stam},
  series    = {London Mathematical Society Lecture Note Series},
  publisher = {Cambridge University Press},
  pages     = {251--291},
  year      = {2021},
  doi       = {10.1017/9781108854207.012}
}

@article{Bree:2026reduce,
  author  = {I. Bree and F. Gasparotto and A. Matija\v{s}i\'c and P. Mazloumi and D. Melnichenko and S. P\"ogel and T. Teschke and X. Wang and S. Weinzierl and K. Wu and X. Xu},
  title   = {New algorithms for {Feynman} integral reduction and $\varepsilon$-factorized differential equations},
  journal = {\href{https://arxiv.org/abs/2511.15381}{arXiv:2511.15381 [hep-th]}},
  year    = {2026}
}

@article{delaCruz:2026seedless,
  author  = {L. de la Cruz and D. A. Kosower},
  title   = {Seedless reduction of {Feynman} integrals},
  journal = {\href{https://arxiv.org/abs/2602.22111}{arXiv:2602.22111 [hep-ph]}},
  year    = {2026}
}

@article{Smirnov:2026fuel,
  author  = {A. V. Smirnov and V. A. Fokin and E. Yu. Chuvashov},
  title   = {Factorization of denominators as a `fuel' for {Feynman} integral reduction},
  journal = {Numer. Methods Program.},
  volume  = {27},
  pages   = {406},
  year    = {2026},
  eprint  = {2605.13256},
  archivePrefix = {arXiv}
}

@article{Li:2026bootstrap,
  author  = {X. Li and D.-M. Mu and Y.-Q. Ma},
  title   = {Analyticity bootstrap of {Feynman} integrals},
  journal = {\href{https://arxiv.org/abs/2610.01154}{arXiv:2610.01154 [hep-ph]}},
  year    = {2026}
}

@article{Zeng:2025rl,
  author  = {M. Zeng},
  title   = {Reinforcement learning and metaheuristics for {Feynman}-integral reduction},
  journal = {Phys. Rev. D},
  volume  = {112},
  pages   = {114051},
  year    = {2025},
  doi     = {10.1103/dmlf-jkfc},
  eprint  = {2504.16045},
  archivePrefix = {arXiv}
}

@article{Song:2026funsearch,
  author  = {Z.-Y. Song and T.-Z. Yang and Q.-H. Cao and M.-X. Luo and H. X. Zhu},
  title   = {Explainable {AI}-assisted optimization for {Feynman} integral reduction},
  journal = {JHEP},
  volume  = {06},
  pages   = {225},
  year    = {2026},
  eprint  = {2502.09544},
  archivePrefix = {arXiv}
}

@article{vonHippel:2025ml,
  author  = {M. von Hippel and M. Wilhelm},
  title   = {Refining integration-by-parts reduction of {Feynman} integrals with machine learning},
  journal = {JHEP},
  volume  = {05},
  pages   = {185},
  year    = {2025},
  doi     = {10.1007/JHEP05(2025)185},
  eprint  = {2502.05121},
  archivePrefix = {arXiv}
}

@article{Berman:2026tube,
  author  = {J. Berman and F. Charton and A. Luna and M. Wilhelm and M. Zeng},
  title   = {Efficient {AI}-inspired reduction of {Feynman} integrals via tube seeding},
  journal = {\href{https://arxiv.org/abs/2606.10698}{arXiv:2606.10698 [hep-ph]}},
  year    = {2026}
}

@article{Huang:2026amflow2,
  author  = {R.-J. Huang and X. Liu and Y.-Q. Ma},
  title   = {{AMFlow} 2.0: significant algorithmic and software improvements for {Feynman} integral evaluation},
  journal = {\href{https://arxiv.org/abs/2607.08477}{arXiv:2607.08477 [hep-ph]}},
  year    = {2026}
}

@article{Britto:2026resonance,
  author  = {R. Britto and T. W. Grimm and A. Hoefnagels},
  title   = {Resonance and differential reduction of {Feynman} integrals},
  journal = {JHEP},
  volume  = {09},
  pages   = {018},
  year    = {2026},
  doi     = {10.1007/JHEP09(2026)018},
  eprint  = {2606.09978},
  archivePrefix = {arXiv}
}

@article{Armadillo:2026review,
  author  = {T. Armadillo},
  title   = {Evaluating {Feynman} integrals through differential equations and series expansions},
  journal = {Eur. Phys. J. Spec. Top.},
  volume  = {234},
  pages   = {7989--8004},
  year    = {2026},
  doi     = {10.1140/epjs/s11734-025-01995-0},
  eprint  = {2502.14742},
  archivePrefix = {arXiv}
}

@article{Liu:2025threeloop,
  author  = {Y. Liu and A. Matija\v{s}i\'c and J. Miczajka and Y. Xu and Y. Xu and Y. Zhang},
  title   = {Analytic computation of three-loop five-point {Feynman} integrals},
  journal = {Phys. Rev. D},
  volume  = {112},
  pages   = {016021},
  year    = {2025},
  eprint  = {2411.18697},
  archivePrefix = {arXiv}
}
